\documentclass[a4paper,12pt]{article}
\usepackage[utf8]{inputenc}
\usepackage{pythontex}
\usepackage{physics}
\AtBeginDocument{\RenewCommandCopy\qty\SI}
\ExplSyntaxOn
\ExplSyntaxOff
\usepackage{algorithm}
\usepackage{algpseudocode}

\usepackage{amsmath}
\usepackage{amssymb}
\usepackage{amsfonts}
\usepackage{graphicx}
\usepackage{hyperref}
\usepackage{natbib}
\usepackage[dvipsnames]{xcolor}
\hypersetup{colorlinks,linkcolor={blue},citecolor={blue},urlcolor={red}} 
\usepackage[margin=1in]{geometry}
\usepackage{booktabs}
\usepackage{multirow}
\usepackage{float}
\usepackage{setspace}
\usepackage{enumitem}
\usepackage{placeins}
\usepackage{flafter}
\usepackage{tikz}
\usepackage[title]{appendix}
\restylefloat{table}
\usepackage{siunitx}

\newtheorem{theorem}{Theorem}

\newtheorem{remark}[theorem]{Remark}

\newtheorem{lemma}[theorem]{Lemma}

\newtheorem{Property}[theorem]{Property}
\newtheorem{corollary}[theorem]{Corollary}

\def\theequation{\thechapter.\arabic{equation}}

\numberwithin{equation}{section}
\numberwithin{theorem}{section}
\newcommand\numberthis{\addtocounter{equation}{1}\tag{\theequation}}

\providecommand{\keywords}[1]
{
  \small	
  \textbf{\textit{Keywords:}} #1
}

\title{Estimation of Piecewise Continuous Regression Function in Finite Dimension using Oblique-axis Regression Tree with Applications in Image Denoising}
\author{\small{Subhasish Basak$^\ast$, Anik Roy$^\dagger$, Partha Sarathi Mukherjee$^\ast$}}

\date{}
\date{\small{\textit{$^\ast$Indian Statistical Institute, Kolkata, India\\
$^\dagger$Emory University, Atlanta, USA}}}

\begin{document}

\maketitle

\begin{abstract}
Decision trees are one of the most widely used nonparametric methods for regression and classification. In existing literature, decision tree-based methods have been used for estimating continuous functions or piecewise-constant functions. However, they are not flexible enough to estimate the complex shapes of jump location curves (JLCs) in two-dimensional regression functions. In this article, we explore the Oblique-axis Regression Tree (ORT) and propose a method to efficiently estimate piece-wise continuous functions in a general finite dimension with fixed design points. The central idea involves clustering the local pixel intensities by recursive tree partitioning and using the local leaf-only averaging for estimation of the regression function at a given pixel. The proposed method can preserve complex shapes of the JLCs well in a finite-dimensional regression function. Due to a different set of assumptions on the underlying regression function, the overall framework of the proofs is different from what is available in the literature on regression trees. Theoretical analysis and numerical results, particularly on image denoising, indicate that the proposed method effectively preserves complicated edge structures while efficiently removing noise from piecewise continuous regression surfaces.
\end{abstract}


\newpage

\section{Introduction}
Tree-based methods hold a pivotal place in modern statistics and machine learning. Classification and Regression Trees (CART) [\cite{breiman1984classification}] are recognized as very powerful and widely used tools among supervised learning algorithms. Over time, ensemble learning techniques that combine multiple trees, such as Random Forest (RF) [\cite{breiman2001random}], bagging, and boosting, have become increasingly popular for classification and regression applications. Decision trees are the building blocks of all these algorithms. The popularity of the techniques based on decision trees is attributed to their intuitive construction and appealing interpretability. However, it has been noticed from the time when CART was first proposed using marginal variables as splitting rules can pose challenges in preserving complex structures of the data while smoothing. This issue can be well addressed by taking a linear combination of the predictors as the splitting rule instead of relying on marginal variables. This procedure is popularly known as the oblique regression tree (ORT) algorithm. In literature, various versions of the ORT algorithms [\cite{cattaneo2024convergence}] are available. However, nearly all of the existing methods assume that the true functional relationship is continuous in nature. However, in practice, there are many situations where the true regression function is clearly discontinuous. Such situations can be recognized when the response variable changes abruptly with small variations in predictor variables. For example, in materials science, rapid phase changes are common during phase transitions. In geostatistics, sensing data from rock strata often exhibits abrupt changes near sediment layers. Similarly, in image analysis, the intensity function of an image changes abruptly around object boundaries or edges. Recently, due to easy availability of various imaging modalities, image monitoring or surveillance has received considerable attention in manufacturing industries, medical science, earth surface surveillance, and so forth. Image denoising is a pivotal pre-processing step in most applications of image monitoring [\cite{roy2024image,roy2025upper}]. As a result, jump-preserving smoothing of regression functions is an extremely important problem with numerous applications across engineering and scientific fields. In this paper, we introduce a nonparametric smoothing technique based on ORT, designed for scenarios where the true regression function is piecewise continuous. Given that a grayscale image intensity function can be represented as a piecewise continuous regression function, we demonstrate the application of the proposed algorithm on the problem of grayscale image denoising in the latter part of the paper.

\subsection{Literature review on related smoothing methods}
So far in the literature, there has been a very limited discussion on the estimation of discontinuous regression functions. Proposed by \cite{breiman1984classification}, CART is one of the most popular nonparametric regression methods that fits the piecewise constant function quite well. However, it fails to capture the complex shape, even if it forms a tree with high depth. A similar nonparametric regression approach in the literature is dyadic regression tree (DRT), proposed by \cite{donoho1997cart}. The central idea is to split the input space at the midpoint along a dimension. The major difference between traditional CART and the DRT is that it allows a recursive splitting at any point of the input space. Although it has computational advantages, and it fits piecewise constant functions reasonably well, it faces similar issues when we are interested in accommodating complex shape boundaries \citep{chaudhuri2023cross}. Recently, ORT-based piecewise continuous function estimation and its theoretical properties have become an active research area \citep{cattaneo2024convergence}. ORT-based decision tree is a natural extension of CART and DRT to accommodate the boundaries of the complex shape. Nowadays, ORT-based decision trees have received remarkable attention among the research communities. \cite{zhan2024consistencyobliquedecisiontree} shows the consistency of the ORT-based algorithm under the assumption that the underlying regression function is continuous. Additionally, several studies in the Gaussian Process (GP) literature are also useful for estimating piecewise continuous regression functions. These methods involve partitioning the input domain into multiple regions and modeling the data in each region with a separate GP model. A Bayesian tree-based GP, proposed by \cite{gramacy2008bayesian}, uses a dyadic treed partitioning to split the input domain along axis-aligned directions, fitting a stationary GP model to the data in each tree leaf. Similarly, \cite{konomi2014adaptive} use a Bayesian CART tree to partition the input domain. However, one major limitation of these piecewise GP models is that their partitioning schemes, such as axis-aligned partitions or triangulations, are too restrictive to capture the complex, curvy boundaries often present in various real-world images.

Since 2D grayscale image surfaces are discontinuous in nature, techniques for estimating piecewise-continuous functions have direct applications in image denoising. A thorough analysis of the history of image denoising can be found in \cite{gonzalez2018digital}. Most of the image denoising methods available in current literature can be grossly classified into two types: (i) Some methods denoise without explicitly estimating underlying edge-structure, whereas (ii) other methods estimate these structures first, and then proceed to denoise. \cite{qiu2005image} provides an elaborated overview of an approach known as Jump Regression Analysis (JRA). 2-D jump regression analysis \citep{qiu2005image} estimates the regression surface from the noisy data, while preserving the boundary of the image object. \cite{qiu2009jump} proposes an edge-preserving image denoising method which splits each neighborhood into two different parts using edge information and uses the best fitting estimates on these regions. However, explicit edge detection-based denoising methods often provide poor results on images with low resolution. To overcome this, \cite{mukherjee20113} develop a clustering-based approach where local neighborhoods around each pixel are split into two clusters using intensity values. There are many other approaches that do not detect the edge structures explicitly. Methods such as bilateral filtering proposed by \cite{chu1998edge} calculate weights using underlying edge information to calculate local averages for smoothing.

\subsection{Contributions of this paper}
As previously discussed, existing literature on JRA offers useful tools for estimating discontinuous functional relationships. However, most of the methods focus on cases with only one or two predictor variables and are not easily extendable to scenarios with multiple predictors. Recently, \cite{kang2021fitting} propose a jump regression model that accommodates multiple predictors, but their approach assumes an additive structure in the functional relationship, effectively disregarding interaction terms among the predictors. To elucidate this limitation, consider the model: $w=f(z_1,z_2)+\varepsilon$ with two predictor variables $z_1$ and $z_2$. Now, the  additive model i.e., $f(z_1,z_2)=f_0+f_1(z_1)+f_2(z_2)$ is suitable for preserving the jump location curves parallel to $X$-axis and $Y$-axis. However, this model is not flexible enough to describe and preserve the jump location curves in other directions. The proposed jump regression model overcomes such limitations of the JRA literature mentioned above. The proposed algorithm does not require pruning for estimating the underlying function; however, we use some constraints to limit the depth of the tree, which may lead to significant computational advantages. Many of the recent tree-based algorithms, e.g., \cite{zhan2024consistencyobliquedecisiontree}, assume the underlying function can be expressed as a \textit{ridge} function. In practice, it is often not a valid assumption. For instance, the image intensity functions are discontinuous in nature and not necessarily a ridge function. In this paper, we do not make any such strong assumptions. Instead, we impose certain assumptions on the underlying discontinuous points to address this issue. The proofs of the theoretical properties of the proposed tree-based method are much simpler, and the overall framework of the proofs is substantially different from what is available in the literature on regression trees. This is due to a different set of assumptions on the underlying regression function. Theoretical properties and their proofs, thus, demonstrate the novelty of this paper, in addition to excellent numerical results.

\subsection{Organization of this paper}
The remainder of this paper is organized as follows. Section \ref{method} describes the proposed methodology. We study its statistical properties in Section \ref{sec::statprop}. Section \ref{application:image denoising} shows one major application of the proposed method in the form of image denoising. Numerical performance of the proposed algorithm is shown in Section \ref{numerical}. A few remarks in Section \ref{conclusion} conclude this paper.

\section{Proposed Methodology} \label{method}
In this paper, we work with the standard regression framework, and the statistical model can be expressed as \vspace{-0.5cm}
\begin{equation}\label{eq:np}
    w(\vb*{z})=f(\vb*{z})+ \varepsilon,
\end{equation}

\vspace{-0.5cm}

\noindent where $w$ is real-valued response or output variable, $\vb*{z}= (z_1,z_2,\ldots z_p)$ is the predictor (input, feature, or covariate vector) variable which lie on an equally space lattice at the points $\Big(\frac{i_1}{n}, \ldots, \frac{i_p}{n}\Big)$, where each $i_j \in {1,2,\dots,n}$ on the design space $\Omega=[0,1]^p$ and, $\varepsilon$ is the error variable. We have a dataset $\mathcal{D}_n= \{(w_i, \vb*{z_i}): 1\leq i \leq n^p\}$, consisting of independent samples drawn from the model in \eqref{eq:np}.  Additionally, we assume that $f$ is \textit{piecewise continuous}, and has the following form:
\begin{equation}
    f(\vb*{z})= \sum_{l=1}^P g_{l}(\vb*{z}) \mathbb{I}_{\Gamma_{l}}(\vb*{z}) 
    \label{eq:jump_model},
\end{equation}


\noindent where $g_l(\vb*{z})=g(\vb*{z})+\delta_l$, and $\delta_l$ is the jump size in $\Gamma_l$. In literature, this is popularly known as jump regression analysis (JRA). Note that, $\{\Gamma_1,\Gamma_2,\ldots,\Gamma_P\}$ is a finite partition of $\Omega$ such that: (i) Each $\Gamma_{l}$ is a connected region in the design space. (ii) Defining $\partial\Gamma_{l}$ as the set of boundary points in $\Gamma_{l}$, $f(\vb*{z})$ is continuous in $\Gamma_{l} \setminus \partial \Gamma_{l}$, for $l= 1,2,\ldots,P$. (iii) $\bigcup_{i=1}^P \Gamma_{l}= \Omega.$ (iv) There exists at most finitely many points $\{\vb*{z}^*,k=1,2,\ldots , K^*\}$ in $[\bigcup_{i=1}^n \partial \Gamma] \bigcap\Omega  $ such that for each $\vb*{z}^*$ with $k=1,2, \ldots K^*$, there are $\Gamma_{k_1}^*,\Gamma_{k_2}^* \in \{\Gamma_{l},{l}=1,2,\ldots P\}$ satisfying 
\begin{enumerate}[label=(\alph*)]
    \item $\vb*{z}^* \in \Gamma_{k_1}^* \bigcap \Gamma_{k_2}^*$, and \vspace{-0.5cm}
    \item $\lim\limits_{\substack{\vb*{z}\rightarrow \vb*{z}^*\\ \vb*{z} \in \Gamma_{k_1}^*}}f(\vb*{z}) = \lim\limits_{\substack{\vb*{z}\rightarrow \vb*{z}^*\\ \vb*{z} \in \Gamma_{k_2}^*}}f(\vb*{z})$.
\end{enumerate}
Condition (a) implies that the point lies on multiple JLCs, whereas condition (b) implies that the limiting value across two partitions created by the JLCs match. In the imaging context, this means that two different partitions of the image are leaking into each other through a point on the relevant JLC. Such continuity points are sporadic, and should be treated differently from the usual continuity points. Our primary objective is to develop a smoothing technique that effectively preserves the jump location curves (JLCs) in the regression surfaces. To achieve this, we partition $\mathcal{D}_n$ in such a manner that each partition contains observations from one side of a JLC. To perform this, we construct a tree using the ORT algorithm, where each leaf node of the tree indicates one partition of the data and contains observations from one side of a JLC. To this end, each partition yields an estimator $\widehat{f}_n$ of $f$ in model \eqref{eq:jump_model} via \textit{leaf-only} averaging.

\begin{subsection}{Recursive partitioning of the data}
In this subsection, we introduce the algorithm by which we construct the recursive ORT, a pivotal step in the context of the proposed jump-preserving surface-smoothing technique. Note that we consider partitioning with respect to hyperplanes only. This restriction is imposed to ensure that the partitions remain convex, which is necessary for the subsequent proofs. However, if we could extend the covariates or pixel coordinates by including nonlinear functions of them, and correspondingly extend the data in a meaningful way, then the results developed here could potentially be applied in that setting to achieve improved performance. Let $\mathcal{N}$ denotes a node of the ORT which is nothing but a subset of $[0,1]^p$, and consider its two children $\mathcal{N}_L= \{\vb*{z}:\vb*{\alpha}^T\vb*{z} \leq c\}$ and $\mathcal{N}_R=\{\vb*{z}:\vb*{\alpha}^T\vb*{z} > c\}$. Note that $\mathcal{N}_L$ and $\mathcal{N}_R$ cannot be empty, and satisfy $\mathcal{N}_L \cup \mathcal{N}_R =\mathcal{N}.$ The central idea behind the tree construction is hierarchically partitioning the data in a greedy manner through a recursive binary splitting rule. Similar to the CART, a daughter node is treated as a parent node in the next step. Then, a parent node $\mathcal{N}$ at any depth is divided into two daughter nodes $\mathcal{N}_L$ and $\mathcal{N}_R$ by maximizing the impurity gain \vspace{0cm}
\begin{equation}\label{eq:impurity_gain}
    \widehat{\Delta}(\mathcal{N},\vb*{\alpha},c)= \frac{1}{|\mathcal{N}|}\bigg[\sum\limits_{\vb*{z} \in \mathcal{N}}\big(w(\vb*{z})-\Bar{w}\big)^2 - \sum\limits_{\vb*{z} \in \mathcal{N}_L}\big(w(\vb*{z})-\Bar{w}_L\big)^2-\sum\limits_{\vb*{z} \in \mathcal{N}_R}\big(w(\vb*{z})-\Bar{w}_R\big)^2 \bigg]
\end{equation}
with respect to $(\vb*{\alpha},c)$. Here, $\Bar{w}$, $\Bar{w}_L$, and $\Bar{w}_R$ are defined as the sample averages of the response variable $w_i$ at the node $\mathcal{N}$, $\mathcal{N}_L$, and $\mathcal{N}_R$, respectively. The impurity gain is nothing but the decrease in the error sum of square (SSE) by increasing the depth of the tree by one step.

Suppose $(\vb*{\widehat{\alpha}},\widehat{c})$ maximizes the Eqn. \eqref{eq:impurity_gain}. We then split the node $\mathcal{N}$ if \vspace{-0.5cm}
$$\widehat{\Delta}(\mathcal{N}) = \widehat{\Delta}(\mathcal{N},\widehat{\vb*{\alpha}},\widehat{c})>r_n,$$

\vspace{-0.5cm}

\noindent where $r_n$ is the splitting threshold for the recursive tree partitioning. Since there is no pruning step in the proposed algorithm, we choose $r_n$ in such a manner that ensures a shallow tree construction. To this end, refinement of $\mathcal{N}$ produces two daughter nodes $\mathcal{N}_L=\{\vb*{z}:\widehat{\vb*{\alpha}}^T\vb*{z} \leq \widehat{c}\}$ and $\mathcal{N}_R=\{\vb*{z}:\widehat{\vb*{\alpha}}^T\vb*{z} > \widehat{c}\}$. These child nodes act as a parent node for the next level of the tree construction, and further recursive refinement carries forward in the similar fashion. At any level of the tree refinement, a node is considered a leaf node if $\widehat{\Delta}(\mathcal{N},\vb*{\alpha},c) \leq r_n$, and we stop the recursion when all nodes become leaf nodes. Finally, we have the leaf nodes that partition the design space, where each partition is nothing but a convex polytope in $[0,1]^p$. Note that the decision tree construction is a filtering step for the proposed jump-preserving smoothing. The intuition behind this is to partition the data in a manner such that the data points on the same side of the JLCs always lie in the same estimated partition. Lemma \ref{lm::dis} in Section \ref{sec::statprop} supports our intuition. A pseudo code outlining the tree construction method is presented in Algorithm \ref{algo_split}.

\noindent {\bf Selection of $r_n$ for numerical implementation:} We choose $r_n$ to be small enough to ensure that there were sufficient partitions to adequately represent the entire image. Note that if we choose this parameter too small, then it will only increase computation time without any improvement in the performance of the estimate. Moreover, $r_n$ should be proportional to the variance of the noise in the image, as that will cause the loss function we used to be proportionally larger. Based on numerical simulations, we choose $r_n = 0.0001\sigma^2$ for all numerical implementations.

\subsection{Proposed jump preserving surface estimator}
Using the proposed algorithm, we divide the design space $[0,1]^p$ in leaf nodes, say $\{\mathcal{L}_n^i | 1\leq i \leq K_n\}$, where $K_n$ denotes the number of leaf nodes, and $\bigcup\limits_{1 \leq i \leq K_n}\mathcal{L}_n^i = [0,1]^p$. For a given point $x\in [0,1]^p$, let $\mathcal{L}_n^{\vb*{x}}$ denote the leaf node where it lies. Then, the proposed estimator is given by:
\begin{equation}\label{estimator}
    \widehat{f}_n(\vb*{x})= \frac{\sum\limits_{z \in \mathcal{L}_n^{\vb*{x}} \bigcap \mathcal{D}_n}w(\vb*z)}{|\mathcal{L}_n^{\vb*{x}}\bigcap \mathcal{D}_n|}.
\end{equation}
From now on, we denote $\mathcal{L}_n\bigcap \mathcal{D}_n$ by $\mathcal{L}_n$ for simplicity. Since the observations in $\mathcal{L}$ lie on one side of a JLC, the proposed estimator preserves the discontinuity well. In the application of image denoising, we slightly modify the proposed estimator. Refer to Eqn. \ref{modified_estimator} for details.

\begin{algorithm} 
\begin{algorithmic}[1]
\caption{Recursive Tree Construction}

\vspace{2mm}

 \State \textbf{Input:} Given dataset $\mathcal{D}_n$ and $r_n$ is the specified cutoff.  
\State $\ell$ is the list of nodes and $\ell_1$ is an empty list. 
\While{$\ell$ is not empty,}
    \State Pick a node $\mathcal{N}$ from $\ell$,
    \State Calculate $\tilde{\Delta} (\mathcal{N}).$
    \If{$\tilde{\Delta} (\mathcal{N}) \leq r_n $}
    \State Remove $\mathcal{N}$ from $\ell$ 
    \State Mark it as a leaf node by adding it to $\ell_1$.
    \Else
    \State Calculate $(\widehat{\vb*{\alpha}}_{opt},\widehat{c}_{opt})=  \arg \max \limits_{(\vb*{\alpha},c)}\widehat{\Delta}(\mathcal{N},\vb*{\alpha},c)$
    \State Split $\mathcal{N}$ into $\mathcal{N}_R$ and $\mathcal{N}_L$ using the line $\{\vb*{\widehat{\alpha}^T_{opt}}\vb*{z} = \widehat{c}\}$.
    \State Remove $\mathcal{N}$ from $\ell$ 
    \State Add $\mathcal{N}_R$ and $\mathcal{N}_L$ to $\ell$.
    \EndIf
    
\EndWhile
\State \textbf{Output:} $\ell_1$ is the list of leaf nodes.
\label{algo_split}
\vspace{2mm}

\end{algorithmic}
   
\end{algorithm}
    
\end{subsection}

\section{Statistical Properties}\label{sec::statprop}
In this section, we investigate certain statistical properties to substantiate our arguments in the proposed methodology. We derive novel theoretical results to establish the consistency of the oblique decision tree within the context of jump regression. Theorems \ref{thm:cardinality} and \ref{thm::fulldim} characterize necessary properties of leaf nodes that should be present in a well-chosen neighborhood for local non-parametric denoising. Lemma \ref{lm::dis} demonstrates that the proposed algorithm correctly distinguishes the continuous regions from the JLCs. Finally, Theorems \ref{thm::asure} and \ref{thm::convrate} ensure asymptotic consistency of $\widehat{f}_n$, and its rate of convergence, respectively. Before going into the details, we introduce relevant notations and assumptions below.

\noindent \textbf{Notations \& Assumptions: } We define a sequence of trees by $\{\mathcal{T}_n\}$, and the sequence of leaf nodes of $\mathcal{T}_n$ by \{$\mathcal{L}_n\}$, where $\mathcal{L}_{(n+1)} \subset \mathcal{L}_n$.  Each successive tree \(\mathcal{T}_{n+1}\) serves as a refinement of \(\mathcal{T}_n\); all internal nodes of \(\mathcal{T}_{n+1}\) are already contained in \(\mathcal{T}_n\), while \(\mathcal{T}_{n+1}\) may include additional leaf nodes corresponding to finer resolution details.
We define $|\mathcal{L}_n|$ as the number of data-points in $\mathcal{D}_n$ which are inside $\mathcal{L}_n$, and define $\mu$ as the Lebesgue measure on $[0,1]^p$. We now make the following assumptions:
\begin{enumerate}[label=A\arabic*:]
    \item $f$ is piecewise Lipschitz continuous.
    \item $f$ is bounded. Without any loss of generality, we assume that it is bounded inside $[0,1]^p$.
    \item The data-points in $\overline{\Lambda_\ell}/int(\Lambda_\ell)$ are called jump points, and the set of jump points has Lebesgue measure zero.
    \item The set of jump location points on each JLC is a piecewise lower-dimensional hyperplane. For notational convenience, each such hyperplane is considered to be a different JLC.
    \item There exists at most countably many JLCs.
    \item The set of singular points has measure zero.
    \item The pointwise noise $\varepsilon$ are independently distributed and follows $\mathbf{N}(0,\sigma^2)$.
    \item The splitting threshold $r_n$ is chosen such that $\sum_n \sqrt{r_n} < \infty$.
\end{enumerate}

\noindent Next, we state a series of important theoretical results regarding the proposed method. All results hold under the assumptions stated above, and proofs appear in Appendix \ref{appen}. The first result simplifies the impurity gain measure.

\begin{Property}
    \begin{equation}
    \hat{\Delta}(\mathcal{N},\vb*{\alpha},c) = \frac{|\mathcal{N}_L||\mathcal{N}_R|}{|\mathcal{N}|^2}\bigg( \Bar{w}_L - \Bar{w}_R \bigg)^2. \label{eq:sim_impurity_gain}
    \end{equation}
\end{Property} 

\noindent The following theorem states that the number of data points in $\mathcal{L}_n$ grows indefinitely as $n$ increases. This result eventually ensures asymptotic consistency of $\widehat{f}_n$.

\begin{theorem}
    Under the assumptions A1-A8,
    $|\mathcal{L}_n|\rightarrow \infty$ almost surely. \label{thm:cardinality}
\end{theorem}
\par
\begin{remark}
    Observe that $\mu(\mathcal{L}_n)$ is non-increasing as $\mathcal{L}_{n+1}$ is a subset of $\mathcal{L}_n$. Since $\mu(\mathcal{L}_n)$ is non-increasing and non-negative, it must have a limit.
\end{remark}

\noindent The following theorem and its corollaries state the results that we can conclude regarding $f$ in both cases when the above limit is non-zero and zero.

\begin{theorem}
Under the assumptions A1-A8, if $\lim\limits_{n\rightarrow\infty}\mu(\mathcal{L}_n) \neq 0$, then $f$ is a.e. constant in $\bigcap\limits_{n}\mathcal{L}_n$. \label{thm::fulldim}
\end{theorem}

\begin{corollary}
     If $\lim\limits_{n\rightarrow\infty}\mu(\mathcal{L}_n) = 0$ and $\bigcap\limits_n\mathcal{L}_n=\mathcal{L}$ lies on a $k$ $(<p)$ dimensional affine subspace of $\mathbb{R}^p$ with $\mu_k(\mathcal{L}) \neq 0$, then one of these is true. \vspace{-0.2cm}
     \begin{enumerate}[label = (\roman*)]
         \item $\mathcal{L}$ contains an entire JLC. \vspace{-0.3cm}
         \item $\mathcal{L}$ is contained entirely inside a JLC. \vspace{-0.3cm}
         \item $f$ is \it{a.e.} constant on $\mathcal{L}$. \vspace{0cm}
     \end{enumerate}\label{cor::kdim}
\end{corollary}

\vspace{-0.5cm}
 
\noindent Using $k=1$ in the above corollary, we get
\begin{corollary}
     Let $\gamma(A)$ denote the maximum distance between two points in the set $A$. If $\lim\limits_{n \rightarrow \infty}\mu(\mathcal{L}_n) = 0$ and $\lim\limits_{n \rightarrow \infty}\gamma(\mathcal{L}_n)\neq 0$, then one of the following is true: \vspace{-0.2cm}
\begin{enumerate}[label=(\roman*)]
\item $\bigcap\limits_n\mathcal{L}_n$ contains an entire JLC. \vspace{-0.3cm}
\item $\bigcap\limits_n\mathcal{L}_n$ is contained entirely inside a JLC. \vspace{-0.3cm}
\item $f$ is \it{a.e.} constant on $\bigcap\limits_n\mathcal{L}_n$.
\end{enumerate} 
\end{corollary}

\begin{Property}
    If $\widehat{\Gamma}$ $=$ $\bigcup\limits_x\Big\{\mathcal{L}^{\vb*{x}}| \mathcal{L}^{\vb*{x}} \textit{ contains an entire JLC and } \mu({\mathcal{L}^{\vb*{x}}})=0\Big\}$, then $\mu(\widehat{\Gamma})=0$. 
\end{Property}
The above property holds because at most countably many sets of these unions are distinctly non-empty, as there are at most countably many JLCs. Next, we show that the average of all continuity points of $f$ over a leaf node converges to the true value. Observe that if $\vb*{x}\in \mathcal{L}_n$ is the point of interest and there is no JLC in $\mathcal{L}_n$, then as $\mu(\mathcal{L}_n) \rightarrow 0$, $\gamma(\mathcal{L}_n)\rightarrow 0$, and $f$ is piecewise Lipschitz, the average of the observed intensity values over the leaf node should converge to $f(\vb*{x})$. However, if $\mathcal{L}_n$ contains a JLC, we need to show that all the points on the leaf node are similar, and do not have any jumps between them, which brings us to the next result.

\begin{lemma}
    Let $\vb*{x}$ be a non-singular continuity point of $f$. If $\lim\limits_{n\rightarrow\infty}\gamma(\mathcal{L}_n^{\vb*{x}})=0$, then $\exists$ $N$ such that $\forall$ $m>N$, $\mathcal{L}_m^{\vb*{x}}$ does not contain any discontinuity point. \label{lm::dis}
\end{lemma}\vspace{-0.1in}

\noindent Finally, the following theorem establishes asymptotic consistency of $\widehat{f}_n$.

\begin{theorem}
Under the assumptions A1-A8,     $\lim\limits_{n\rightarrow \infty}\widehat{f}_n(\vb*{x}) =f(\vb*{x})$ \textit{a.e.}, almost surely. \label{thm::asure}
\end{theorem}

\noindent The next theorem provides a convergence rate of $\widehat{f}_n$ for points with non-zero derivatives.

\begin{theorem}
Let $\vb*{x}_0$ be a point such that $f'$ is continuous at $\vb*{x}_0$, and $\lvert f'(\vb*{x}_0) \rvert > 2\delta$ where $\delta > 0$. This ensures the existence of a neighborhood 
$N_{\vb*{x}_0}$ of $\vb*{x}_0$ such that, for all $\vb*{x} \in N_{\vb*{x}_0}$,
$ \lvert f'(\vb*{x}) - f'(\vb*{x}_0) \rvert < \delta \quad \text{and} \quad \lvert f'(\vb*{x}) \rvert > \delta.$
Then, under the assumptions A1-A8, the proposed estimator $\widehat{f}_n(\vb*{x})$ as in \eqref{estimator} converges to $f(\vb*{x}_0)$ at the rate $o\!\left(\Big(\sqrt{\,r_n  \Big| \log \big| \log r_n \big| \Big|\, } \, C_{\chi^2_1}\Big)^{1/4}\right)$, $C_{\chi^2_1}$ being a constant depending on a $\chi^2_1$ random variable.
\label{thm::convrate}
\end{theorem}

\section{An Application of the Proposed Method: Image Denoising} \label{application:image denoising}
Image denoising is a fundamental problem in the field of image processing. It serves as a crucial pre-processing step in many applications to make further analysis meaningful and reliable. For example, in sequential image surveillance, denoising plays a central role. In addition to removing noise, a key requirement for effective denoising methods is the preservation of edge structures within the image.
In the JRA literature [\cite{qiu2005image}], a 2D grayscale image is expressed as a discontinuous regression surface, where the edges of objects in the image are treated as points of discontinuity or ``jump points." In this section, we introduce a new image denoising method based on an oblique-axis regression tree, designed to preserve these jumps in the regression surface. Although the algorithm is versatile and applicable to various types of images, we focus on 2D grayscale images here for simplicity. The proposed 2D grayscale image denoising technique directly applies the algorithm from Section 2 with $p=2$. Consequently, the 2D JRA model can be described similarly as follows: \vspace{-0.5cm}
\begin{equation}
    w_{ij}= f(x_i,y_j)+ \varepsilon_{ij} \hspace{0.1cm}, \quad \mbox{for} \quad i,j=1,2, \ldots, n,
\end{equation}

\begin{figure}[htbp!]
\centering
\includegraphics[height=3cm,width=10.5cm]{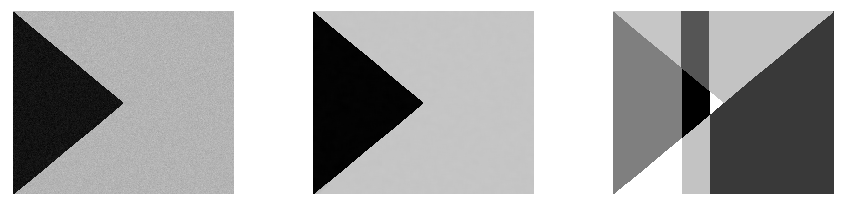}
\caption{Demonstration of partitioning of the pixels in the test image. Images from left to right are the noisy image, the denoised image by the proposed method, and estimated partitions, respectively. The partitions are represented by different shades. Note that the estimated partitions are subsets of the true partitions.}\label{fig:demo}
\end{figure}

\vspace{-0.5cm}

\noindent where $\mathcal{D}_n=\{(x_i,y_j):i,j=1,2,....,n\}$ are equally spaced design points (or pixels) in the design space $\Omega=[0,1] \times [0,1]$, $f(x,y)$ is the unknown image intensity function at $(x,y)$, and $\varepsilon_{ij}$ are independent and identically distributed (i.i.d.) random errors with mean $0$ and variance $\sigma^2 > 0$. To denoise an image, we perform recursive tree splitting until all nodes become leaf nodes, for a specified cut-off $r_n$ depending on $n$. Using the tree partitioning method outlined in Algorithm \ref{algo_split}, we partition the design space $[0,1]^2$ in $K_n$ number of convex polygons, represented as $\{\mathcal{L}_n^i | 1\leq i \leq K_n\}$. The resulting estimator is obtained by averaging the image intensity values within the leaf nodes. However, in practice, large leaf nodes may cause the loss of intricate details in the image surface. To address this issue, we consider \textit{leaf-wise local weighted} averaging instead of averaging over the whole leaf node. For this adjustment, we consider a square neighborhood at each pixel $(x,y) $ as $B(x,y;h_n)$ where $2h_n >0$ is the length of each side of the square. The modified estimator at $f(x,y)$ is thus defined as:
\begin{equation}\label{modified_estimator}
    \widehat{f}^M_n(x,y)= \frac{\sum\limits_{(u_i,v_j) \in \mathcal{L}_n^{(x,y)}\bigcap B(x,y;h_n)}w(u_i,v_j)SS_{(x,y)}(u_i,v_j)}{\sum\limits_{(u_i,v_j) \in \mathcal{L}_n^{(x,y)}\bigcap B(x,y;h_n)}SS_{(x,y)}(u_i,v_j)}.
\end{equation}
Here, $\mathcal{L}_n^{(x,y)}$ represents the leaf node containing the test pixel coordinate $(x,y)$, $(u_i,v_j)$ denotes pixel coordinates within $\mathcal{L}_n^{(x,y)}$, and $SS_{(x,y)}(u_i,v_j)$ is a similarity score between the pixels $(x,y)$ and $(u_i,v_j)$ expressed as follows:
\[
SS_{(x,y)}(u_i,v_j) = \exp\bigg(-\frac{\kappa_n||B(u_i,v_j;h_n)-B(x,y;h_n)||^2}{(nh_n)^2}\bigg).
\]
In the above expression, the distance between the neighborhoods represents the $L_2$-distance between the pixel intensities.
Regarding the choice of $h_n$, selecting a value too large could blur fine details, while a very small value may result in a noisy estimator $\widehat{f}^M_n(x,y)$. Based on numerical experience, we recommend choosing $h_n \in [\frac{2}{n},\frac{4}{n}]$. Note that the proposed method is not very sensitive to the choices of $h_n$. For numerical implementation, we select $h_n = \frac{3}{n}$, and the scaling parameter $\kappa_n = 5$. This is based on checking with various values of $h_n$ and $\kappa_n$ on the various images. Figure \ref{fig:demo} demonstrates that the proposed method can partition the pixels of a given image appropriately, and thus can remove noise while preserving the edge details.


\section{Numerical Studies}\label{numerical}
In this section, we evaluate the performance of the proposed method through numerical analysis, comparing it with several state-of-the-art methods from the literature. Since jump preserving surface estimation is common in image analysis, we conduct our numerical experiments with various simulated images and real images. Including the noise removal, another major aspect in the context of image denoising is edge preservation. We demonstrate the performance of the proposed method both in terms of noise removal and edge preservation. To evaluate the noise removal performance of the proposed method in comparison with the state-of-the-art methods, we consider root-expected mean square error (REMSE), defined as 
$$REMSE(\widehat{f},f)= \sqrt{\mathbb{E}\bigg[\sum\limits_{(u,v) \in \Omega}\bigg( \widehat{f}(u,v)-f(u,v)\bigg)^2\bigg]}.$$
On the other hand, the performance of edge preserving can be quantified by a similarity measure between the sets of detected edge pixels of the denoised image and the true image. One such similarity measurement between the denoised image and the true image can be expressed as \vspace{-0.2cm}
\[
d_{KQ}(\widehat{\Gamma}_{\widehat{f}}, \widehat{\Gamma}_{{f}})= \frac{0.5}{|\widehat{\Gamma}_{\widehat{f}}|} \sum_{(u,v) \in \widehat{\Gamma}_{\widehat{f}}}d_E((u,v), \widehat{\Gamma}_{{f}})+\frac{0.5}{|\widehat{\Gamma}_{{f}}|} \sum_{(u,v) \in \widehat{\Gamma}_{{f}}}d_E((u,v), \Gamma_{\widehat{f}}),
\]

\vspace{-0.2cm}

\noindent where $\widehat{\Gamma}_{\widehat{f}}$ and $\widehat{\Gamma}_{f}$ are the detected edge pixels for the denoised image and the true image, respectively. It is to be noted that $d_{KQ}$ does not hold the properties of a metric. One popular metric to compute the distance between two point sets is \textit{Hausdorff} distance. However, \textit{Hausdorff} metric is very sensitive to individual points. Therefore, we select $d_{KQ}$ to compute the similarity measurement between the point-sets of the detected edge pixels of the two images. For a good image denoising procedure, it is expected that both the values of REMSE and $d_{KQ}$ would be small.  Edge preservation can be assessed with any edge-detection method.
In our numerical studies, we adopt the local linear kernel (LLK) approach available in the DRIP package [\cite{drip}] on CRAN-R, particularly due to its robustness in the presence of noise.

\subsection{A brief description of the competing methods}
To evaluate the numerical performance of the proposed method in terms of REMSE and jump preservation, we select the following three popular state-of-the-art methods: (i) image denoising using usual random forest technique [\cite{breiman2001random}], (ii) edge preserving image denoising method proposed by \cite{qiu2009jump}, and (iii) image denoising method by non-local means proposed by \cite{buades2011non}.

\textbf{(i) Image denoising using random forest (RF):} Random forest is the most popular tree-based ensemble learning algorithm in the context of nonparametric regression. Since it is an ensemble method, it is capable of accommodating the complex shapes in the data. One of the major advantages of the random forest over a single tree-based method (CART) is that it can avoid the problem of overfitting, unlike CART. Therefore, random forest could be used as a potential image denoising method. However, in the case of image smoothing, to accommodate the complex structure of the image object, random forest splits more, leading to deeper trees, which aggravates the problem of overfitting. The proposed method also involves a decision tree, and therefore it is comparable with random forest-based methods.

\textbf{(ii) Jump preserving local linear kernel smoothing (JPLLK):} The method first divides the local neighborhood into two parts using gradient information. Then, three different estimates are calculated using local linear kernel regression using the left, right, and whole neighborhood, respectively. The best of these three locally fitted surfaces is chosen, depending on their mean squared error. Then, we follow the same procedure on the fitted surface again. But this time, one of those three estimates is chosen as the final estimate based on their estimated variances. This method is very good at preserving jumps, although its image denoising performance falls short when compared to our proposed method.

\textbf{(iii) Image denoising based on non-local means (NLM):} In this method, the true image intensity value at a given pixel is estimated using a weighted average on a large portion of the entire image. To calculate the weights, each pixel is scored according to its similarity with the given pixel. Then, these scores are used as weights for the estimation. For convenience, we choose an implementation proposed by \cite{froment2014parameter}, as it automatically calculates the parameters for the denoising method. Among these competing methods, the non-local means method usually performs at best at image denoising in the sense of visual appearances, at the cost of blurring the image.

\subsection{Simulations}
\begin{figure}[ht!]
    \centering
    \includegraphics[height=4cm,width=16cm]{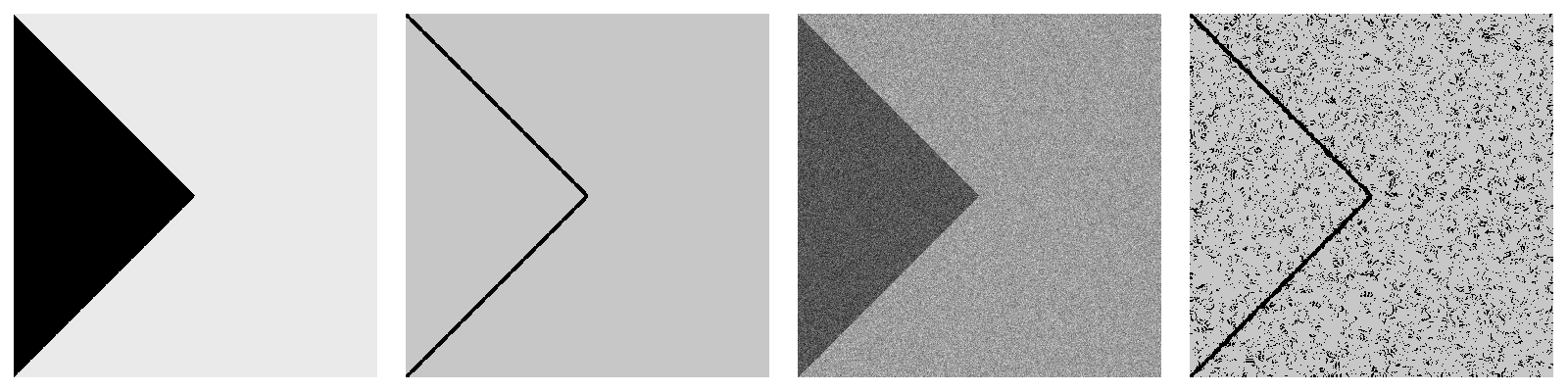}
    \includegraphics[height=4cm,width=16cm]{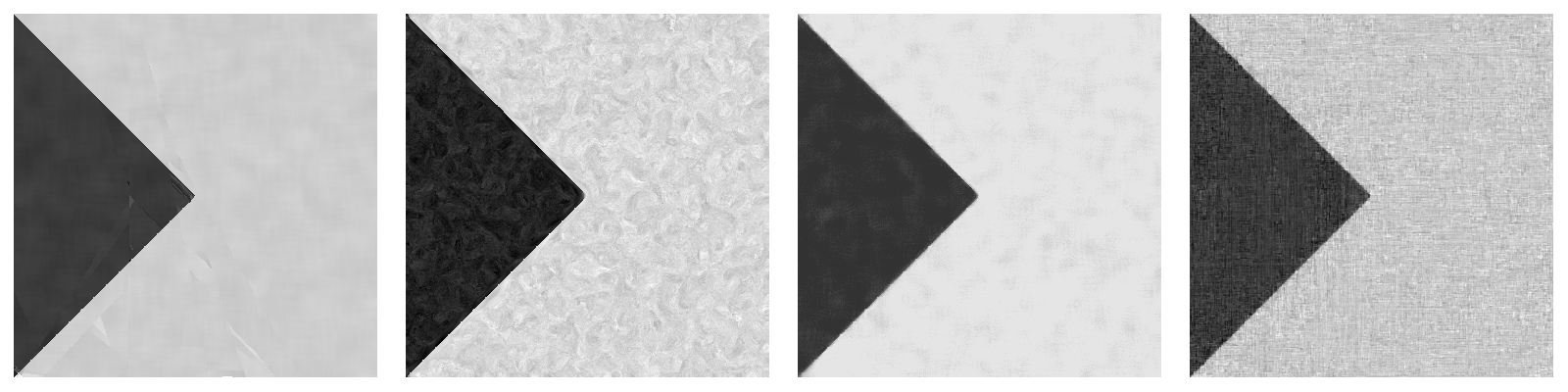}
    \includegraphics[height=4cm,width=16cm]{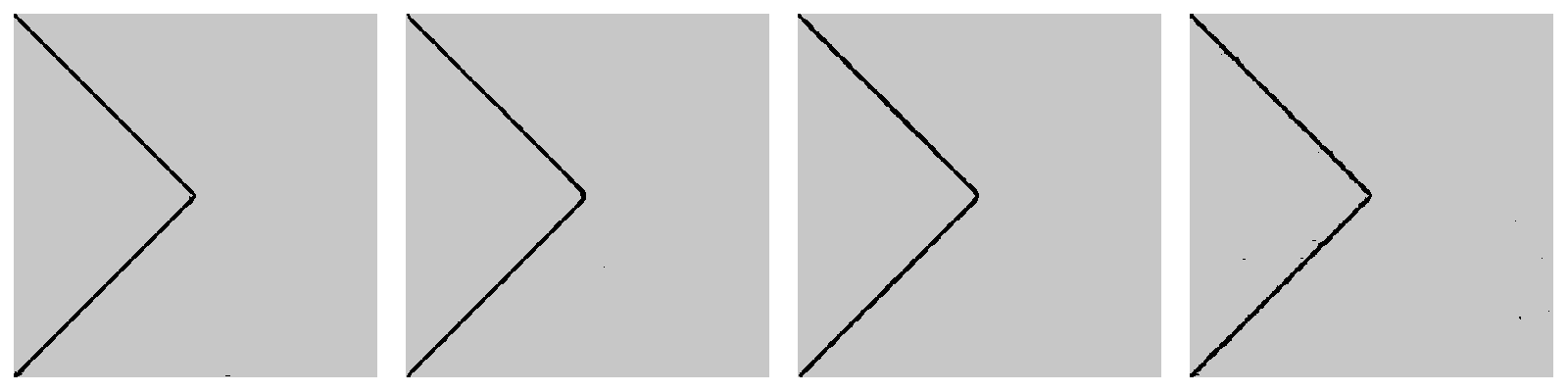}
    \caption{Performance comparison on the simulated test image with point-wise Gaussian noise with $\sigma=0.3$. The first row shows the original image, original edges, noisy image, and detected edges on the noisy image, respectively. The second row shows denoised images produced by the proposed method, JPLLK, NLM, and RF. The third row shows the detected edges of the corresponding denoised images.}
     \label{fig:simulated}
\end{figure}

In the simulation study, we consider a triangle image as the test image. See the extreme left of the upper panel in Figure \ref{fig:simulated} for the true triangle image. The image intensity function of the test image can be expressed as 
\begin{equation}
    f(u,v)= \mathbb{I}[(u,v) \in \Omega_1],\quad (u,v) \in [0,1] \times [0,1],
\end{equation}
where $\mathbb{I}$ is the indicator function, and $\Omega_1$ denotes the region enclosed by the triangle. Throughout this section, we use Gaussian additive noise to generate the noisy image surface. See the third image from the left of the upper panel in Figure \ref{fig:simulated} for the noisy test image.
We carry out the simulation study under the following settings: (i) to demonstrate the noise removal property, we use three different noise levels with $\sigma=0.1, 0.2, \ \text{and} \ 0.3$. (ii) to demonstrate the consistency of the proposed estimation method, we perform further simulations with two different image resolutions:  $200\times 200$, i.e., $n=200$, and $400\times 400$, i.e., $n=400$.
Table \ref{tab:sim_mse_performance} shows the REMSE values for various cases and methods. Moreover, the summary of the performances of the concerned methods with respect to edge preservation is provided in Table \ref{tab:sim_edge_performance}. Based on 100 replications, the results are shown in Table \ref{tab:sim_mse_performance}.

\begin{table}[htp]
\footnotesize
    \centering
    \begin{tabular}{ c|c|c|ccc }
    & & \multicolumn{1}{c|}{Proposed Method} & \multicolumn{3}{c}{Competing Methods} \\
    \hline
     Resolution & Noise & ORT & JPLLK & NLM   & RF\\
    \hline
    \multirow{3}{4em}{$200 \times 200$}    &  0.3 & 33.8 (4.697)   & 54.1 (1.196) & 51.1 (1.559) & 108.3 (1.095)   \\
    & 0.2  & 25.5 (4.237)   & 40.0 (0.659)  & 29.6 (1.136) & 86.5 (0.728) \\
    & 0.1  & 18.1 (4.752)   & 27.4 (0.382)  & 14.4 (0.511) & 70.6 (0.717) \\
    \hline 
    \multirow{3}{4em}{$400 \times 400$}    &  0.3 & 22.3 (4.024) & 49.5 (0.539) & 40.0 (0.821) & 91.6 (0.852)    \\
    & 0.2 &
    15.4 (3.822)   & 33.9 (0.318)  & 24.9 (0.584) & 69.6 (0.522) \\
    & 0.1 & 7.9 (3.349)  & 19.8 (0.203)  & 12.5 (0.255) & 53.0 (0.470)  \\
\end{tabular}
    \caption{Comparisons of various methods on the simulated test image using (REMSE$\times 10^3$) values based on $100$ independent replications with standard error within the parentheses}.
    \label{tab:sim_mse_performance}
\end{table}

\begin{table}[ht]
\footnotesize
    \centering
    \begin{tabular}{ c|c|c|ccc }
    & & \multicolumn{1}{c|}{Proposed Method} & \multicolumn{3}{c}{Competing Methods} \\
    \hline
     Resolution & Noise & ORT  & JPLLK & NLM & RF \\
    \hline
    \multirow{3}{4em}{$200 \times 200$}    &  0.3 & 0.427    & 0.649 & 0.722 & 2059 \\
    & 0.2  & 0.115    & 0.265  & 0.151 & 0.470 \\
    & 0.1  & 0.110    & 0.266  & 0.000 & 0.391\\
    \hline 
    \multirow{3}{4em}{$400 \times 400$}    &  0.3 & 0.086    & 0.570 & 0.386 & 1.67\\
    & 0.2 & 0.086   & 0.087  & 0.059 &0.275 \\
    & 0.1 & 0.005   & 0.068  & 0.003 & 0.177\\
\end{tabular}
    \caption{Comparisons of various methods regarding jump preservation on the simulated test image using $(d_{KQ} \times 10^3)$.}
    \label{tab:sim_edge_performance}
\end{table}
From Tables \ref{tab:sim_mse_performance} and \ref{tab:sim_edge_performance}, we can see that the proposed ORT-based method performs better when $n=400$ rather than $200$, in terms of both REMSE and edge preservation. As the proposed method is asymptotically consistent, these results are intuitively reasonable. The proposed method is uniformly better than RF and JPLLK both in noise removal and edge preservation. Except for the case when $n=200$ and  $\sigma=0.1$, the proposed method outperforms NLM in all other cases. The denoised images are shown in the middle panel of  Figure \ref{fig:simulated}. From the visual impression, it appears that the denoising performances of JPLLK and RF are relatively poor compared to the proposed method. NLM performs relatively better; however, several noisy patchy regions are present in the denoised image. We demonstrate the edge detection performances in the lower panel of Figure \ref{fig:simulated}. From this simulation study, we see that the proposed method outperforms the competing methods in almost all scenarios considered above.
\begin{figure}[ht!]
    \centering
    \includegraphics[height=4cm,width=16cm]{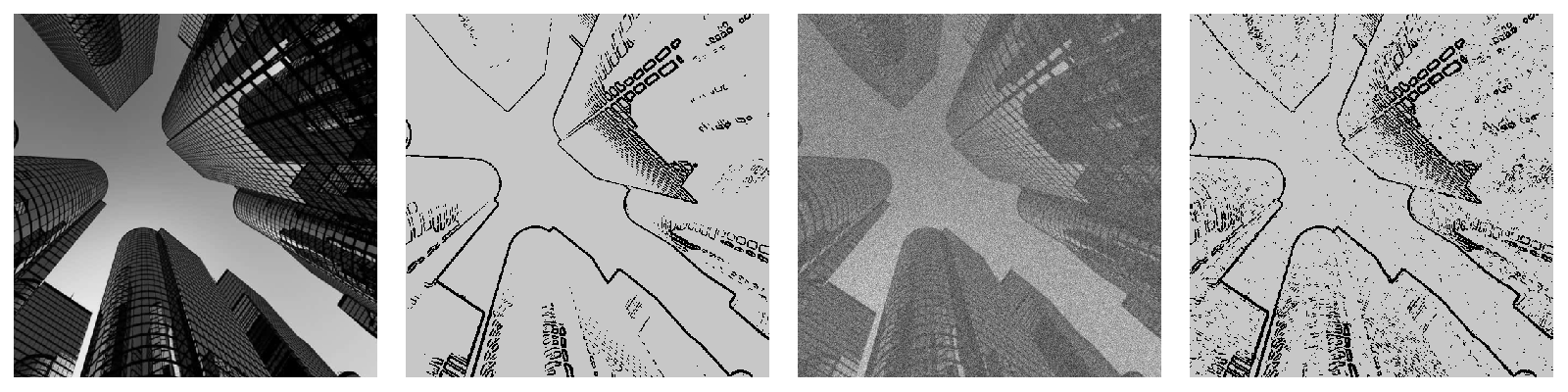}
    \includegraphics[height=4cm,width=16cm]{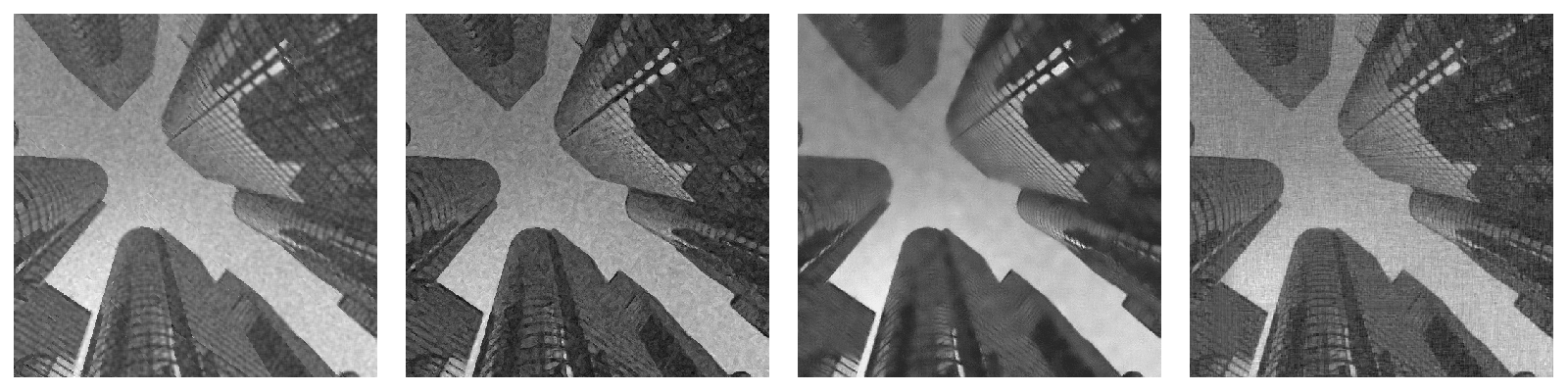}
    \includegraphics[height=4cm,width=16cm]{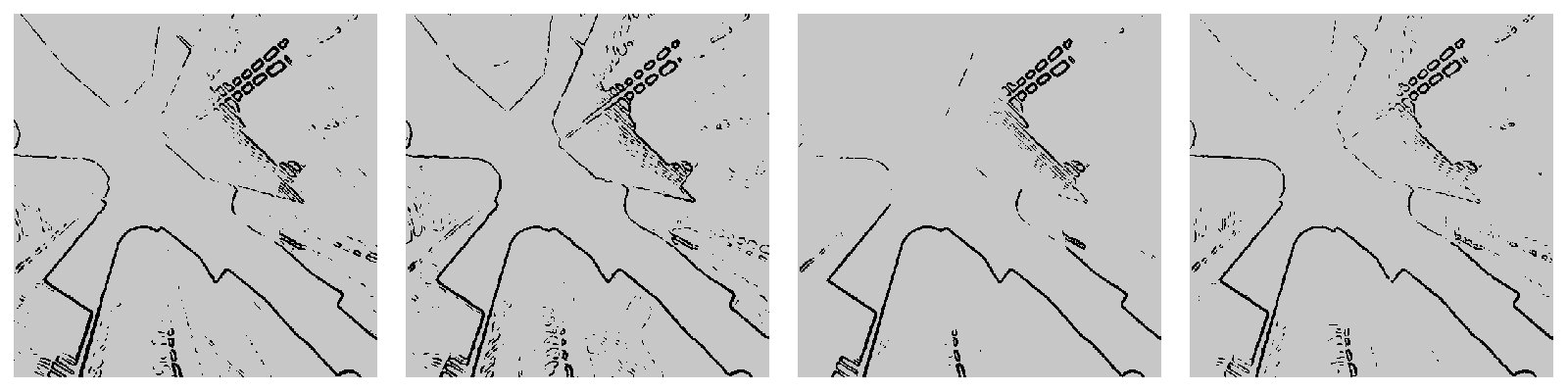}
    
    \caption{Performance comparison on the building image with point-wise Gaussian noise with $\sigma=0.2$. The first row shows the original image, original edges, noisy image, and detected edges on the noisy image, respectively. The second row shows denoised images produced by the proposed method, JPLLK, NLM, and RF. The third row shows the detected edges of the corresponding denoised images.}
     \label{fig:bui}
\end{figure}

\subsection{Applications on real images}
\begin{figure}[ht!]
    \centering
    \includegraphics[height=4cm,width=16cm]{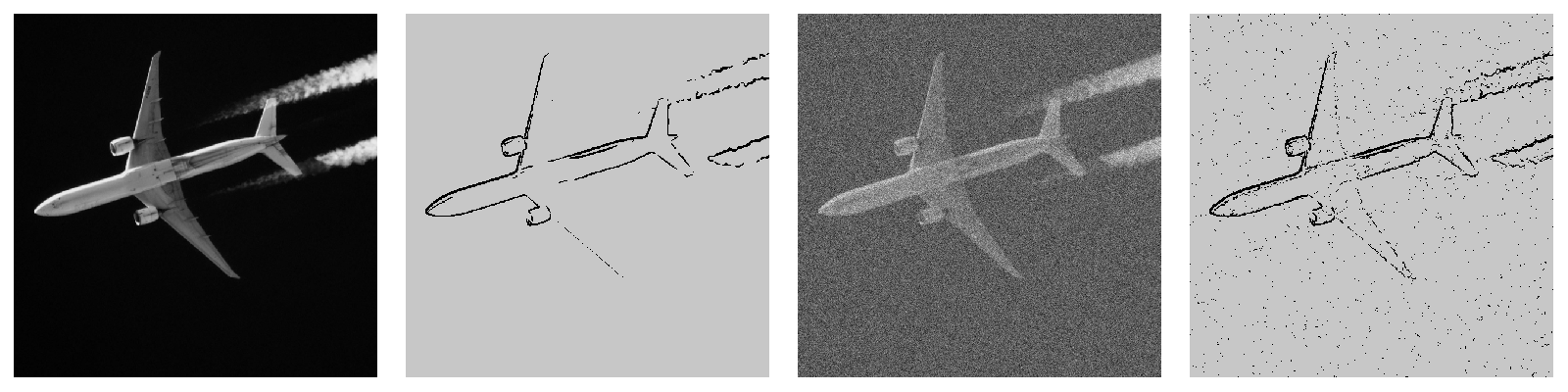}
    \includegraphics[height=4cm,width=16cm]{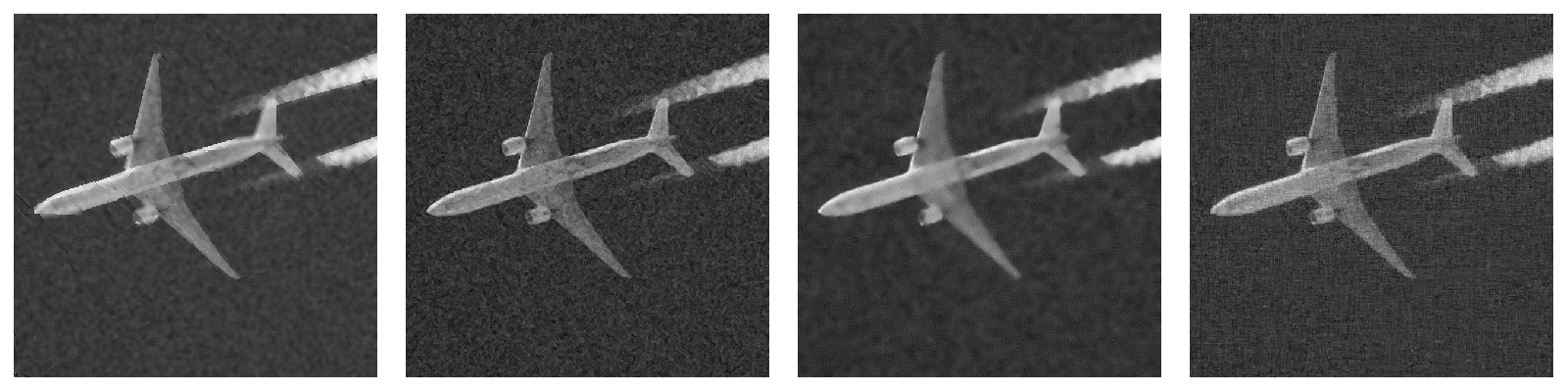}
    \includegraphics[height=4cm,width=16cm]{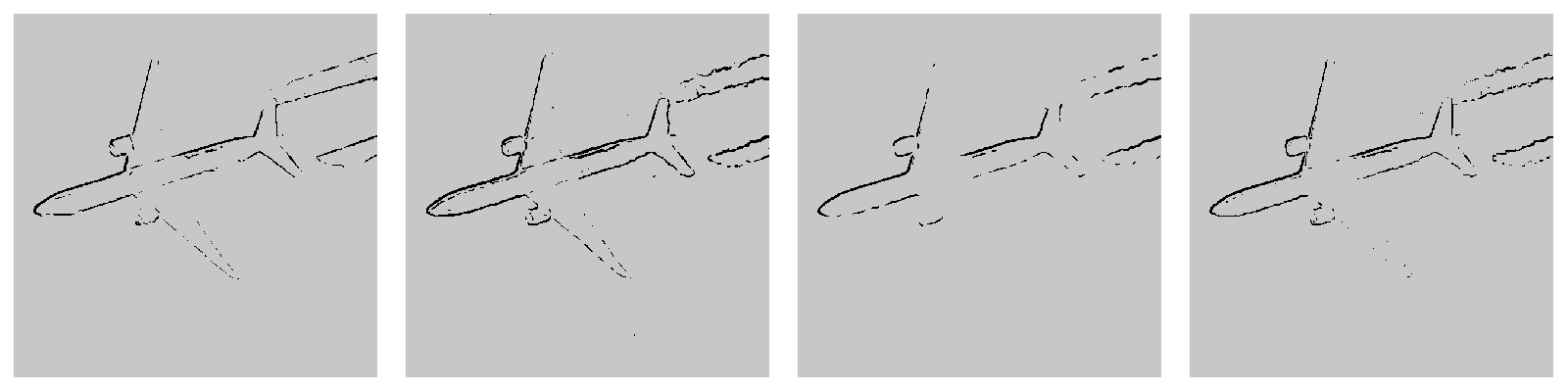}
    \caption{Performance comparison on the aero-plane image with point-wise Gaussian noise with $\sigma=0.2$. The first row shows the original image, original edges, noisy image, and detected edges on the noisy image, respectively. The second row shows denoised images produced by the proposed method, JPLLK, NLM, and RF. The third row shows the detected edges of the corresponding denoised images.}
     \label{fig:plane}
\end{figure}

We apply the proposed ORT-based method on complicated real images and perform a comparative analysis with the state-of-the-art competing methods. In this regard, we consider two different real images. The image in the extreme left of the upper panel of Figure \ref{fig:bui} shows high-rise buildings with resolution $523 \times 523$. The image in the extreme left of the upper panel of Figure \ref{fig:plane} shows a moving aero-plane with resolution $628 \times 628$.
The numerical performances of the concerned methods regarding noise removal and edge preservation are provided in Tables \ref{tab:real_mse_perform} and \ref{tab:real_edge_performance}. From Table \ref{tab:real_mse_perform}, we see that for the building image, the proposed method uniformly outperforms JPLLK and RF in terms of REMSE. In this regard, the NLM method is comparatively better than the proposed method except for larger noise levels, when its performance is similar to the proposed method. In terms of edge preservation, the proposed method is much better than NLM in most cases. Although edge-preservation measures are good for RF and JPLLK, their performance on noise removal is rather poor, especially when the noise level is high. In summary, we can conclude that the performance of the proposed method shows a reasonable trade-off between noise removal and edge preservation abilities. The proposed method has good image denoising capability along with a good edge preservation property.
\begin{table}[ht]
\footnotesize
    \centering
    \begin{tabular}{ c|c|c|ccc }
    & & \multicolumn{1}{c|}{Proposed Method} & \multicolumn{3}{c}{Competing Methods} \\
    \hline
     Image & Noise & ORT & JPLLK & NLM   & RF\\
    \hline
    \multirow{3}{4em}{Building}    &  0.3 & 95.8 (0.873)    & 102.2 (0.324) & 96.2 (0.343)& 106.1 (0.497)  \\
    & 0.2  & 82.1 (0.402)   & 100.2 (0.221)  & 77.0 (0.248)  & 84.8 (0.415) \\
    & 0.1  & 58.8 (3.22)  & 80.3 (0.153) & 48.6 (0.142) & 64.1 (0.269)  \\
    \hline 
    \multirow{3}{4em}{Plane}    &  0.3 & 45.4 (0.741)    & 89.0 (0.505) & 44.6 (0.301) & 82.1 (0.829) \\
    & 0.2 & 37.4 (0.621)  & 60.9 (0.183)  & 33.2 (0.191)  & 56.1 (0.507) \\
    & 0.1 & 27.7 (0.704)   & 33.9 (0.099)  & 16.7 (0.082) & 34.8 (0.299)\\
\end{tabular}
    \caption{Comparisons of various methods on the real images using (REMSE$\times 10^3$) values based on $100$ independent replications with standard error within the parentheses.}
    \label{tab:real_mse_perform}
\end{table}
\begin{table}[ht]
\footnotesize
    \centering
    \begin{tabular}{ c|c|c|ccc }
    & & \multicolumn{1}{c|}{Proposed Method} & \multicolumn{3}{c}{Competing Methods} \\
    \hline
     Image & Noise & ORT & JPLLK & NLM   & RF\\
    \hline
    \multirow{3}{4em}{Building}    &  0.3 & 13.2   & 4.14 & 65.16 & 8.53  \\
    & 0.2  & 6.86   & 4.14  & 25.26 & 6.67 \\
    & 0.1  & 1.59   & 1.37  & 2.38 & 5.63 \\
    \hline 
    \multirow{3}{4em}{Plane}    &  0.3 & 8.34    & 15.25 & 31.45 & 3.46 \\
    & 0.2 & 4.56   & 2.33  & 6.51 & 1.67 \\
    & 0.1 & 1.63   & 1.10 & 0.75 & 2.29 \\
\end{tabular}
    \caption{Comparisons of various methods regarding jump preservation on real images using $(d_{KQ} \times 10^3)$.}
    \label{tab:real_edge_performance}
\end{table}
\FloatBarrier

\subsection{Comparison with a state-of-the-art deep learning model}
We now compare the proposed algorithm with one of the best-performing state-of-the-art deep learning based models. Note that this is not an apple-to-apple comparison, as the deep learning based models are trained on multiple images of various types, while the proposed method is for a single image only. The model named {\it Restormer}, proposed by \cite{zamir2022restormer}, is a transformer-based model tailored for image restoration tasks such as denoising. Unlike conventional convolutional approaches, it employs multi-head attention and feeds the forward network to effectively model both complex local textures and long-range spatial dependencies. This design allows {\it Restormer} to suppress noise while preserving structural details and natural image fidelity. As a result, it achieves state-of-the-art performance in real image denoising and other restoration problems. Table \ref{tab:restormer_mse_perform} shows the performance of the proposed algorithm when compared to the pre-trained model {\it Restormer} under the same setup as before. This is reasonable because the images on which {\it Restormer} was originally trained are possibly different from the test images we apply to.
\begin{table}[ht]
\footnotesize
    \centering
    \begin{tabular}{ c|c|c|c }
    & & \multicolumn{1}{c|}{Proposed Method} & {Competing Method} \\
    \hline
     Image & Noise & ORT & Restormer\\
    \hline
    \multirow{3}{4em}{Building}    &  0.3 & 95.8 (0.873)    & 165.7(0.5)  \\
    & 0.2  & 82.1 (0.402)   & 100.2(0.3) \\
    & 0.1  & 58.8 (3.22)  & 52.7(0.1)  \\
    \hline 
    \multirow{3}{4em}{Plane}    &  0.3 & 45.4 (0.741)    & 154.1(0.4) \\
    & 0.2 & 37.4 (0.621)  & 75.2(0.4) \\
    & 0.1 & 27.7 (0.704)   &35.4(0.1)\\
\end{tabular}
    \caption{Comparisons with {\it Restormer} on the real images using (REMSE$\times 10^3$) values based on $100$ independent replications with standard error within the parenthesis.}
    \label{tab:restormer_mse_perform}
\end{table}
From Table \ref{tab:restormer_mse_perform}, we can see that the proposed algorithm works much better in high noise scenarios when compared to {\it Restormer}. Note that the performance of {\it Restormer} may be affected by the high noise used in the demonstrated cases, and it may not have been trained using noise of this magnitude. For implementation of {\it Restormer}, we utilize the official codes and pre-trained weights provided by \cite{zamir2022restormer}.

\subsection{Performance comparison on a large-scale dataset}
We further evaluate the proposed method using the SIDD+NTIRE challenge dataset \citep{inproceedings}. For this analysis, the validation dataset is employed, as it provides publicly available ground-truth images. The dataset comprises 32 base images, each with 32 cropped variants, resulting in a total of 1024 images with a spatial resolution of 256$\times$256. In the NTIRE challenge, 22 participating teams submitted 24 distinct algorithms, all trained on the validation data and subsequently assessed on the test set. Our proposed algorithm attains a peak signal-to-noise ratio, i.e., PSNR of $33.291$ on the sRGB benchmark. This result is noteworthy, as it situates the proposed algorithm within the competitive range of the leading methods, including some of the best-performing deep learning approaches reported in the original challenge. Considering that the proposed method does not rely on training with large datasets, it also demonstrates broader applicability and potential for generalization.

\subsection{Applications on 3-D image denoising}
\begin{center}
    \begin{figure}[ht!]
        \centering
        \includegraphics[width=16cm,height=12cm]{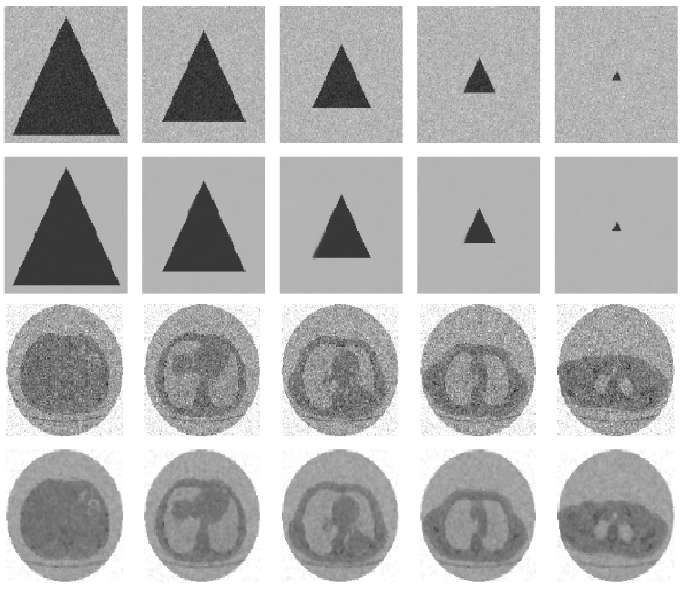}
        \caption{The top two rows show various slices of the noisy and denoised tetrahedron image, while the bottom two rows show the corresponding results for a real lung CT image. The columns correspond to the slices along the Z-axis at indices 13, 38, 64, 90, and 115.}
        \label{fig:3dimage}
    \end{figure}
\end{center}

In this section, we demonstrate the application of the proposed algorithm to 3-D image denoising. In a simulation study, we consider a tetrahedron with its base aligned parallel to the XY-plane. We then add Gaussian noise with various standard deviations to the image for testing the denoising performance of the proposed method. The simulated image has a resolution of $128 \times 128 \times 128$. In the real-data setting, we use a lung CT scan from the dataset provided by \cite{LungPETCTDx2020}. The image is rescaled to $128 \times 128 \times 128$, intensities are normalized to $[0,1]$, and then point-wise Gaussian noise is applied. We evaluate the performance of ORT under noise levels with standard deviations of $0.1$, $0.2$, and $0.3$ across $10$ independent trials. The results of ORT on both simulated and real images are presented in Figure~\ref{fig:3dimage} and Table~\ref{tab:3d_image}.
\begin{table}[ht]
\footnotesize
    \centering
    \begin{tabular}{c|c|c|c}
        Noise Level & $\sigma=0.1$ & $\sigma=0.2$ & $\sigma=0.3$\\
        \hline
        Simulated Image & 0.0140 (0.0053) & 0.0267(0.0046)  & 0.03831 (0.0047) \\
        Real Image & 0.0394  (0.0003)    & 0.0536  (0.0002)& 0.0702 (0.0001)\\
    \end{tabular}
    \caption{Performance evaluation of the proposed method for 3-D image denoising using RMSE.}
    \label{tab:3d_image}
\end{table}

\section{Concluding Remarks} \label{conclusion}
In this paper, we propose a framework using Oblique-axis Regression Tree to estimate the underlying discontinuous function on a lattice and show its point-wise convergence with an explicit convergence rate under mild assumptions. We also show that similar points are often grouped in the same leaf node of the decision tree we obtain from the proposed algorithm, which is very desirable since it facilitates further analysis. Then, we proceed to apply this algorithm to noisy images for removing noise, and show its superior performance compared to many other state-of-the-art methods. However, there are scopes for improvement. In this paper, we only show that the algorithm converges when the number of lattice points increases to infinity. Instead, one can consider a different approach and keep the number of lattice points fixed while letting the number of samples go to infinity. In this way, one can extend the use of the proposed algorithm to various real-life problems, such as image monitoring. The partitioning is restricted to linear types only to ensure that the partitions remain convex, which is necessary for subsequent proofs. However, if we could extend the covariates or pixel coordinates by including nonlinear functions of them, and correspondingly extend the image in a meaningful way, then the results developed here could potentially be applied in that setting to achieve improved performance. Instead of working with equally spaced design points, one can expand its capability to work with general tabular data, where the design points are not necessarily evenly distributed. Another future direction of research is to generate multiple sample images using bootstrapping techniques and build a random forest using the proposed algorithm to facilitate better estimates. Apart from these, the proposed algorithm can also be applied to other problems, e.g., image segmentation, denoising of an image sequence, and so on.

\begin{appendices}

\section{}\label{appen}
Below, we provide the proofs of the theoretical results.
\subsection{Proof of Property 3.1}
From {\bf Eqn. 2.3}, we have
\begin{align*}
    & \widehat{\Delta}(\mathcal{N},\vb*{\alpha},c) \allowdisplaybreaks\\
    &= \frac{1}{|\mathcal{N}|}\bigg(\sum\limits_{\vb*{z} \in \mathcal{N}} w(\vb*z)^2 - \sum\limits_{\vb*{z} \in \mathcal{N}_L} w(\vb*z)^2 - \sum\limits_{\vb*{z} \in \mathcal{N}_R} w(\vb*z)^2 -|\mathcal{N}|\Bar{w}^2+|\mathcal{N}_R|\Bar{w}_R^2+|\mathcal{N}_L|\Bar{w}_L^2\bigg) \allowdisplaybreaks\\
    &= \frac{1}{|\mathcal{N}|}\bigg( |\mathcal{N}_R|\Bar{w}_R^2+|\mathcal{N}_L|\Bar{w}_L^2 -|\mathcal{N}|(\frac{|\mathcal{N}_R|\Bar{w}_R}{|\mathcal{N}|}+\frac{|\mathcal{N}_L|\Bar{w}_L}{|\mathcal{N}|})^2 \bigg) \allowdisplaybreaks\\
    &=\frac{1}{|\mathcal{N}|^2}\bigg( (|\mathcal{N}||\mathcal{N}_R|-|\mathcal{N}_R|^2)\Bar{w}_R^2+(|\mathcal{N}||\mathcal{N}_L|-|\mathcal{N}_L|^2)\Bar{w}_L^2 - 2|\mathcal{N}_R||\mathcal{N}_L|\Bar{w}_L\Bar{w}_R \bigg) \allowdisplaybreaks\\
    &= \frac{|\mathcal{N}_L||\mathcal{N}_R|}{|\mathcal{N}|^2}\bigg( \Bar{w}_L - \Bar{w}_R \bigg)^2.
\end{align*}

\subsection{Proof of Theorem 3.2}
Since $\mathcal{L}_n$ are leaf nodes, we have $\widehat{\Delta}(\mathcal{L}_n)\leq r_n$.

\noindent Now, consider any $(\vb*{\alpha},c)$ that splits $\mathcal{L}_n$ in half. Then, using {\bf Eqn. 3.5} for that split, we have:
$r_n \geq \frac{|\mathcal{L}_n^L||\mathcal{L}_n^R|}{|\mathcal{L}_n|^2}\bigg( \Bar{w}_L - \Bar{w}_R \bigg)^2$. We know $\frac{|\mathcal{L}_n^L||\mathcal{L}_n^R|}{|\mathcal{L}_n|}\bigg( \Bar{w}_L - \Bar{w}_R \bigg)^2$ follows noncentral $\chi^2_1(k)$ with a non-centrality parameter which we call $k$. Therefore, we have
\[
    r_n \geq \frac{\chi^2_1(k)}{|\mathcal{L}_n|} \implies |\mathcal{L}_n| \geq \frac{\chi^2_1(k)}{r_n}.
\]

\vspace{0cm}

\noindent As $\chi^2_1(k)$ is stocahastically larger than $\chi^2_1$, we can say: \vspace{-0.5cm}
\[
\mathbb{P}(|\mathcal{L}_n| \leq M) \leq \mathbb{P}(\chi^2_1\leq Mr_n).
\]

\vspace{-0.5cm}

\noindent As $r_n \rightarrow 0$, we can conclude that $|\mathcal{L}_n| \rightarrow \infty$ in probability.
If we choose $r_n$ such that the RHS of the above inequality becomes summable, then we will have almost sure convergence as well. In that case, we use a bound given by \cite{ghosh2021exponential}:
\begin{align*}
    \mathbb{P}(\chi^2_1\leq Mr_n) &\leq exp\bigg[\frac{1}{2}\bigg(1-Mr_n+log(Mr_n)\bigg)\bigg] \\
                                &= \sqrt{Mr_n}exp\bigg(\frac{1-Mr_n}{2}\bigg) \leq \sqrt{eMr_n}.
\end{align*}

\vspace{-0.5cm}

\noindent Therefore, if $\{\sqrt{r_n}\}$ is summable, we have $|\mathcal{L}_n| \rightarrow \infty$ almost surely.

\subsection{Proof of Theorem 3.4}
Suppose, $\mu(\mathcal{L}_n)\rightarrow \ell$. Since $|\mathcal{L}_n| \approx n^p\mu(\mathcal{L}_n)$, for large $n$, we also have $|\mathcal{L}_n|\rightarrow \infty$.
Assume $\bigcap\limits_{n}\mathcal{L}_n= \mathcal{L}$. As $\mu(\mathcal{L}) = \ell > 0$, we have $\mathcal{L}_n\rightarrow \mathcal{L}$ which is non empty.  Now consider a split along the direction $\vb*{\alpha}$ and constant $c$, which splits $\mathcal{L}$ into $\mathcal{L}^L$ and $\mathcal{L}^R$. The same hyperplane also splits $\mathcal{L}_n$ in $\mathcal{L}_n^L$ and $\mathcal{L}_n^R$. Then, we must have $\mathcal{L}_n^L\rightarrow \mathcal{L}^L$ and $\mathcal{L}_n^R\rightarrow \mathcal{L}^R$. As $\mathcal{L}_n$ are leaf nodes, we must have $\widehat{\Delta}(\mathcal{L}_n,\vb*{\alpha},c) \leq r_n$. Therefore,
\begin{align*}
         \frac{1}{|\mathcal{L}_n|}\bigg[\sum\limits_{\vb*{z} \in \mathcal{L}_n}\big(w(x,y)-\Bar{w}\big)^2 -& \sum\limits_{\vb*{z} \in \mathcal{L}_n^R}\big(w(x,y)-\Bar{w}_R\big)^2-\sum\limits_{\vb*{z} \in \mathcal{L}_n^L}\big(w(x,y)-\Bar{w}_L\big)^2 \bigg] \leq r_n \\
         \implies \frac{|\mathcal{L}_n^L||\mathcal{L}_n^R|}{|\mathcal{L}_n|^2}\bigg(\Bar{w}_L - \Bar{w}_R\bigg)^2 &\leq r_n.\\
         \implies \frac{\mu(\mathcal{L}_n^L)\mu(\mathcal{L}_n^R)}{\mu(\mathcal{L}_n)^2}\bigg(\Bar{w}_L - \Bar{w}_R&\bigg)^2 \leq r_n.\\ 
\end{align*}

\vspace{-1cm}

\noindent Note that $\Bar{w}_R$ and $\Bar{w}_L$ are the averages of right and left child respectively. Now observe: \vspace{-1cm}
 \begin{align*}
     \Bar{w}_L  & = \sum\limits_{\vb*{z} \in \mathcal{L}_n^L}w(\vb*{z})/|\mathcal{L}_n^L| \\
                & = \sum\limits_{\vb*{z} \in \mathcal{L}_n^L}(f(\vb*{z})+\epsilon(\vb*{z}))/|\mathcal{L}_n^L| \\
                &= \frac{\frac{1}{n^p}\sum\limits_{\vb*{z} \in \mathcal{L}_n^L}f(\vb*{z})}{\frac{|\mathcal{L}_n^L|}{n^p}} +\frac{\sum\limits_{\vb*{z} \in \mathcal{L}_n^L}\epsilon(\vb*{z})}{|\mathcal{L}_n^L|}\\
                & \rightarrow \bigg(\int_{\mathcal{L}^L} f(\vb*{z})d\mu\bigg)/\mu(\mathcal{L}^L).
 \end{align*}   

\vspace{0cm}

\noindent As the first term is the Riemann sum of the integral of $f(\vb*{z})\mathbb{I}_{\mathcal{L}_n^L}(\vb*{z})$, which is bounded by $f$, we can use Dominated Convergence Theorem (DCT) to get that limit. The second term converges to $0$, due to the Strong Law of Large Numbers (SLLN).
As $r_n \rightarrow 0$, combining the above two results, we have:
\begin{equation}
 \frac{\mu(\mathcal{L}^L)\mu(\mathcal{L}^R)}{\mu(\mathcal{L})^2}\bigg(\frac{\int_{\mathcal{L}^L} f(\vb*{z})d\mu}{\mu(\mathcal{L}^L)} - \frac{\int_{\mathcal{L}^R} f(\vb*{z})d\mu}{\mu(\mathcal{L}^R)}\bigg)^2 = 0. \label{eq::1}
\end{equation}
As $\mu(\mathcal{L}) \neq 0$, and the choices of $\alpha$ and c were arbitrary, we note that:
 \begin{align}
\frac{\int_{\mathcal{L}^L} f(\vb*{z})d\mu}{\mu(\mathcal{L}^L)}&= \frac{\int_{\mathcal{L}^R} f(\vb*{z})d\mu}{\mu(\mathcal{L}^R)} \nonumber \\
\implies\frac{\int_{\mathcal{L}^L} f(\vb*{z})d\mu}{\mu(\mathcal{L}^L)}&=\frac{\int_{\mathcal{L}} f(\vb*{z})d\mu}{\mu(\mathcal{L})} \nonumber\\
\implies \frac{\int_{\vb*{\alpha}^Tz \leq c} f(\vb*{z})d\mu}{\mu(\{\vb*{\alpha}^Tz \leq c\}\bigcap \mathcal{L})} &=\frac{\int_{\mathcal{L}} f(\vb*{z})d\mu}{\mu(\mathcal{L})} \label{eq::CW}
\end{align}
for all possible splits. Define $h(\vb*{z})=\frac{f(\vb*{z})}{\int_{\mathcal{L}} f(\vb*{z})d\mu}$. Consider a random variable $\vb*{X}$ which takes values inside $\mathcal{L}$ with density $h(z)$. From Eqn.  \eqref{eq::CW}, we see that:
\[
\int_{\vb*{\alpha}^Tz \leq c}h(\vb*{z})d\mu = \frac{\mu(\{\vb*{\alpha}^Tz \leq c\}\bigcap \mathcal{L})}{\mu(\mathcal{L})}
= \mathbb{P}(\vb*{\alpha}^T\vb*{X} \leq c) .
\]

\noindent Therefore, all linear combinations of $\vb*{X}$ have the same distribution as linear combinations of a uniform random variable on $\mathcal{L}$. Hence, by Cram\'er-Wold theorem, we can say $\vb*{X}$ has uniform distribution on $\mathcal{L}$, i.e., $h(z)=\frac{1}{\mu(\mathcal{L})}\text{ \it a.e.} \implies f(\vb*{z}) = \frac{\int_{\mathcal{L}} f(\vb*{z})d\mu}{\mu(\mathcal{L})}\text{ \it a.e}$. Hence $f(\vb*{z})$ is \textit{a.e.} constant on $\mathcal{L}$.
\par
\vspace{20pt}

\subsection{Proof of Corollary 3.5}
As $\mathcal{L}_n$ are convex, so is $\mathcal{L}$. Therefore, if $\mu(\mathcal{L})=0$, then it is a convex set of dimension less than $p$. Suppose, the dimension of $\mathcal{L}$ is $k <p$. Consider $\mu_k$ to be the Lebesgue measure corresponding to $\mathbb{R}^k$ on $\mathcal{L}$. Hence, we have $\mu_k(\mathcal{L})\neq 0$.

Now as $\mathcal{L}$ is of dimension $k$, there must exist $(n-k)$ axes which are not parallel to the k-dimensional
affine subspace $\Bar{L}$ which contains  $\mathcal{L}$. Call them $e_1,e_2, \dots e_{n-k}$. Let $\mathcal{R}_{n-k}$ be the subspace generated by $e_1,e_2, \dots e_{n-k}$. Now consider 
$\mathcal{R}_x = \vb*{z_x} + \mathcal{R}_{n-k}$, where $z_x$ is the closest lattice point to $\vb*{x}$, and define a new function: \vspace{-0.5cm}
$$J_n(\vb*{x}) =  \frac{\sum\limits_{y\in \mathcal{R}_x\bigcap \mathcal{L}_n}w(\vb*{y})}{|\mathcal{R}_x\bigcap \mathcal{L}_n|} \hspace{10pt} \forall \vb*{x}\in \Bar{L}. $$

\vspace{0cm}

\noindent We are basically superimposing the whole image onto $\Bar{L}$. Now, if we define $\Delta_J(\mathcal{L}_n)$ in a similar fashion to $\Delta$, it is easy to observe that $\Delta_J(\mathcal{L}_n) \leq \Delta(\mathcal{L}_n)\leq r_n$.
Proceeding similarly to the proof of the above theorem, we get: \vspace{-0.2cm}
\begin{align*}
\frac{\mu_k(\Bar{\mathcal{L}}_n^L)\mu_k(\Bar{\mathcal{L}}_n^R)}{\mu(\Bar{\mathcal{L}}_n)^2}\bigg(\Bar{J}_n^L - \Bar{L}_n^R\bigg)^2 &\leq r_n,
\end{align*}

\vspace{-0.2cm}

\noindent where \vspace{0cm}
\begin{align*}
\Bar{J}_n^L &= \sum\limits_{\vb*{x} \in \Bar{\mathcal{L}}_n^L}J_n(\vb*{x})/ | \Bar{\mathcal{L}}_n^L|\\
          &= \int\limits_{\Bar{\mathcal{L}}_n^L}J_n(\vb*{x})d\mu_k{x}/(| \Bar{\mathcal{L}}_n^L|/n^k),
\end{align*}
and $\Bar{J}_n^R$ is defined in a similar fashion. Observe that if $f$ is continuous on $\mathcal{L}$, $J_n(\vb*{x})$ converges point-wise to $f(x)$. Using dominated convergence theorem (DCT) on the integral, we get: \vspace{-0.5cm}
\[
    \int\limits_{\Bar{\mathcal{L}}_n^L}J_n(\vb*{x})d\mu_k{x} \rightarrow \int\limits_{\mathcal{L}^L}f(\vb*{x})d\mu_k{x}.
\]

\vspace{-0.5cm}

\noindent Then, proceeding similarly as the proof of {\bf Theorem 3.4}, we get $f$ is a.e. constant in $\mathcal{L}$.
Remember that for small enough $\mu(\mathcal{L}_n)$, we assume only one JLC is present in $\mathcal{L}_n$. Hence, if $f$ is not continuous on a measure set with respect to $\mu_k$ on $\mathcal{L}$, the same result holds.
Otherwise, because of the assumption of a JLC being approximately a lower-dimensional plane, and $\mathcal{L}$ being convex, it must mean that the JLC contains $\mathcal{L}$ entirely or $\mathcal{L}$ contains the JLC entirely.

\subsection{Proof of Corollary 3.6}
Applying $k=1$ in {\bf Corollary 3.5}.

\subsection{Proof of Property 3.7}
Property 4 holds because at most countably many sets of these unions are distinctly non-empty, as there are at most countably many JLCs.

\subsection{Proof of Lemma 3.8}
Suppose that for all $n$, $\mathcal{L}_n^{\vb*{x}}$ contains at least one discontinuity point. Consider $\mathcal{N}_\epsilon(\vb*{x})$ to be an $\epsilon$ ball around $\vb*{x}$. 
As $\gamma(\mathcal{L}_n^{\vb*{x}})\rightarrow 0$, we can say that for every $\epsilon$, there exists an $n_\epsilon > 0$ such that $\forall \hspace{2pt} m > n_\epsilon $, $\gamma(\mathcal{L}_n^{\vb*{x}})<\epsilon$, which implies $\mathcal{L}_m^{\vb*{x}} \subseteq \mathcal{N}_\epsilon(\vb*{x})$.
However, as $\mathcal{L}_m^{\vb*{x}}$ contains at least one discontinuity point, say $\vb*{y_n}$. Since $\{\vb*{y_n}\}$ is a bounded sequence, there exists a convergent subsequence. Assume that the convergent subsequence converges to $\vb*{y}$. As $\vb*{y}\in \overline{\bigcap\limits_n\mathcal{L}_n^{\vb*{x}}}$, we have $\norm{\vb*{y-x}}=0 \implies \vb*{y=x}$. Since the set of jump points is closed by definition, either $\vb*{y}$ is a discontinuity point or a singular point. Both options contradict our assumption that $\vb*{x}$ is a non-singular continuity point. Therefore, $\mathcal{L}_n^{\vb*{x}}$ cannot contain a discontinuity point $\forall$ $n$. Since $\mathcal{L}_n^{\vb*{x}}$ is a non-increasing sequence of sets, we conclude that after some $N$, it does not contain any discontinuity point.

\subsection{Proof of Theorem 3.9}
We prove this only on continuity points for which $\mathcal{L}^{\vb*{x}}$ \textit{does not contain an entire JLC} as the remaining points have measure zero. We divide this proof into three different cases:

\noindent\textbf{Case 1:} $f$ is not \textit{a.e.} constant in $\bigcap\limits_{n}\mathcal{L}_n^{\vb*{x}}$. Hence we also have $\lim\limits_{n\rightarrow\infty}\mu(\mathcal{L}_n^{\vb*{x}}) = 0$.\\
As per the assumptions, $\bigcap\limits_{n}\mathcal{L}_n^{\vb*{x}}$ cannot lie entirely on a JLC. Hence, we have $\lim\limits_{n\rightarrow\infty}\gamma(\mathcal{L}_n^{\vb*{x}})$. Using {\bf Lemma 3.8}, $\exists$ $N$ such that $\mathcal{L}_n^{\vb*{x}}$ does not contain any JLC  $\forall$ $n > N$. In this case, $\forall \vb*{z} \in \mathcal{L}_n^{\vb*{x}}$, we have: \vspace{-0.5cm}
\begin{align*}
    \abs{w(\vb*{z})-f(\vb*{x})} &\leq \abs{f(\vb*{z})-f(\vb*{x})} + \abs{\varepsilon(\vb*{z})} \leq C_\ell\gamma((\mathcal{L}_n^{\vb*{x}})) + \abs{\varepsilon(\vb*{z})}\\
    \implies \abs{\sum\limits_{z \in \mathcal{L}_n^{\vb*{x}}}w(\vb*z)-f(x)} &\leq 
    |\mathcal{L}_n^{\vb*{x}}|C_\ell\gamma(\mathcal{L}_n^{\vb*{x}})+ \abs{\sum\limits_{z \in \mathcal{L}_n^{\vb*{x}}}\varepsilon(\vb*{z})} \\
    \implies \abs{\widehat{f}_n(\vb*{x})-f(x)} &\leq C_\ell\gamma(\mathcal{L}_n^{\vb*{x}}) + \abs{\frac{\sum\limits_{z \in \mathcal{L}_n^{\vb*{x}}}\varepsilon(\vb*{z})}{|\mathcal{L}_N^{\vb*{x}}|}} \numberthis \label{eq::temp2} \\
    \implies \abs{\widehat{f}_n(\vb*{x})-f(x)} &\xrightarrow{n\rightarrow \infty} 0, \text{ using SLLN and {\bf Corollary 3.5}}.
\end{align*}

\vspace{-0.5cm}

\noindent \textbf{Case 2:} $f$ is \textit{a.e.} constant in $\bigcap\limits_{n}\mathcal{L}_n^{\vb*{x}}$ and $\lim\limits_{n\rightarrow\infty}\mu(\mathcal{L}_n^{\vb*{x}}) \neq 0$.

\noindent In this case, we know that $f$ is \textit{a.e.} constant in $\bigcap\limits_{n}\mathcal{L}_n^{\vb*{x}}$. So we have: \vspace{-0.5cm}
\begin{align*}
    {\widehat{f}_n(\vb*{x})-f(x)} &= \frac{\sum\limits_{z \in \mathcal{L}_n^{\vb*{x}}}(f(\vb*{z})-f(\vb*{x}))}{|\mathcal{L}_n^{\vb*{x}}|} + \frac{\sum\limits_{z \in \mathcal{L}_n^{\vb*{x}}}\varepsilon(\vb*{z})}{|\mathcal{L}_n^{\vb*{x}}|}\\
    &= \frac{\frac{1}{n^p}\sum\limits_{z \in \mathcal{L}_n^{\vb*{x}}}(f(\vb*{z})-f(\vb*{x}))}{|\mathcal{L}_n^{\vb*{x}}|/n^p} + \frac{\sum\limits_{z \in \mathcal{L}_n^{\vb*{x}}}\varepsilon(\vb*{z})}{|\mathcal{L}_n^{\vb*{x}}|}.
\end{align*}

\vspace{-0.5cm}

\noindent The numerator of the first term converges to $\int\limits_{\bigcap\limits_n\mathcal{L}_n^{\vb*{x}}}(f(\vb*{z})-f(\vb*{x}))$ which is 0 as f is almost surely \textit{a.e.} constant there, and the denominator converges to $\lim\limits_{n\rightarrow\infty}\mu(\mathcal{L}_n^{\vb*{x}})$ which is non-zero. Therefore, we have $\lim\limits_{n\rightarrow \infty}\widehat{f}_n(\vb*{x}) =f(\vb*{x})$ a.s.

\vspace{0.25cm}

\noindent \textbf{Case 3:} $f$ is \textit{a.e.} constant on $\bigcap\limits_{n}\mathcal{L}_n^{\vb*{x}}$ and $\lim\limits_{n\rightarrow\infty}\mu(\mathcal{L}_n^{\vb*{x}}) = 0$. \\
Observe that if $\exists$ $k$ such that $\lim\limits_{n\rightarrow\infty}\mu_k(\mathcal{L}_n^{\vb*{x}}) \neq 0$, we can proceed similar to Case 2. Therefore, without any loss of generality, we can assume $\gamma(\mathcal{L}_n^{\vb*{x}})\rightarrow 0$, and then proceed similarly to Case 1.

\noindent From above cases, we conclude that $\lim\limits_{n\rightarrow \infty}\widehat{f}_n(\vb*{x}) =f(\vb*{x})$ \textit{a.e.}, almost surely.

\subsection{Proof of Theorem 3.10}
While the following diagram of $\mathcal{L}_n$ is drawn for the 2-D case, the general $p$-dimensional situation can be understood similarly. Note that the actual leaf node is a convex polygon in the 2-D case and a convex polyhedron in the general $p$-dimensional scenario. For the simplicity of drawing, the shape is shown as an ellipse rather than a convex polygon with many sides.
\begin{center}
\begin{tikzpicture}[scale=0.7]
  \def\a{4}   
  \def\b{2.5} 

  \draw[thick] (0,0) ellipse [x radius=\a, y radius=\b];

  \coordinate (A) at ({\a*cos(20)},  {\b*sin(20)});
  \coordinate (B) at ({\a*cos(120)}, {\b*sin(120)});
  \coordinate (C) at ({\a*cos(215)}, {\b*sin(215)});
  \coordinate (D) at ({\a*cos(310)}, {\b*sin(310)});

  \draw[very thick] (A)--(B)--(C)--(D)--cycle;

  \draw[dashed, thick] (A)--(C);
  \draw[dashed, thick] (B)--(D);

  \foreach \P/\pos in {A/above right, B/above left, C/below left, D/below right}{
    \fill (\P) circle (2pt) node[\pos] {\P};
  }
\end{tikzpicture}

{\footnotesize {\bf Diagram showing the BD-axis and AC-hyperplane of a leaf node $\mathcal{L}_n$.}}
\end{center}
The line $BD$ represents the maximum possible distance in the set along a given direction $\vb*{u}$. $AC$ is perpendicular to the gradient $f^{'}(\vb*{x}_0)$, and hence, for the general $p$-dimension, it is a hyperplane. Next, we want to split the line $BD$ along $AC$. In the general situation, we can split it using the plane $(f^{'}(\vb*{x}_0))^t(\vb*{x}-\vb*{x}_0)=0$. The intuition behind this is as follows. As the gradient along $AC$ is close to zero, we can ignore changes in $f(x)$ along the line BD. Moreover, we can move the line or hyperplane in such a manner that it splits the set into two equal parts, which lets us equate the fraction $\frac{n_1n_2}{|\mathcal{L}_n|^2}$ to $\frac{1}{4}$.

Without any loss of generality, we can assume that $\vb*{x_0}$ is the point of intersection. If it is not, we can work with the intersection point itself, but the value of the function $f$ at that point will be similar to the given point, as $|f^{'}(\vb*{x})|$ lies between $3\delta$ and $\delta$. We then denote the two children of the leaf node $\mathcal{L}$ by $A$ and $B$. Approximating the average intensity by the corresponding Riemann integral and applying the SLLN to the random errors, we have
\begin{align*}
    \bar{w}_1 & \approx \frac{\int\limits_Af(\vb*{x})d\vb*{x}}{\mu(A)} + O\left(\sqrt{\frac{\log\log|\mathcal{L}_n|}{|\mathcal{L}_n|}}\right) \allowdisplaybreaks \\
              & = \frac{1}{\mu(A)}\int\limits_A\Big(f(\vb*{x_0})+(f^{'}(\vb*{\zeta})^t(\vb*{x}-\vb*{x_0})\Big)d\vb*{x} + O\left(\sqrt{\frac{\log\log|\mathcal{L}_n|}{|\mathcal{L}_n|}}\right) \allowdisplaybreaks\\
              & = \frac{1}{\mu(A)}\int\limits_A\Big(f(\vb*{x_0})+(f^{'}(\vb*{x_0})+\vb*{v})^t(\vb*{x}-\vb*{x_0})\Big)d\vb*{x} + O\left(\sqrt{\frac{\log\log|\mathcal{L}_n|}{|\mathcal{L}_n|}}\right) \allowdisplaybreaks\\
              & = \frac{1}{\mu(A)}\int\limits_A\Big(f(\vb*{x_0})+(f^{'}(\vb*{x_0}))^t(\vb*{x}-\vb*{x_0})\Big)d\vb*{x} + O\Big(|\vb*{v}|\gamma(\mathcal{L}_n)\Big) + O\left(\sqrt{\frac{\log\log|\mathcal{L}_n|}{|\mathcal{L}_n|}}\right) \allowdisplaybreaks\\
              & =f(\vb*{x_0}) + \frac{1}{\mu(A)}\int\limits_A(f^{'}(\vb*{x_0}))^t(\vb*{x}-\vb*{x_0})d\vb*{x} + O\Big(|\vb*{v}|\gamma(\mathcal{L}_n)\Big) + O\left(\sqrt{\frac{\log\log|\mathcal{L}_n|}{|\mathcal{L}_n|}}\right) \allowdisplaybreaks\\
              &=f(\vb*{x_0}) + \frac{1}{\mu(A)}\int\limits_A(f^{'}(\vb*{x_0}))^t(\vb*{x}-\vb*{x_0})d\vb*{x} + O(|\vb*{v}|) + O\left(\sqrt{\frac{\log\log|\mathcal{L}_n|}{|\mathcal{L}_n|}}\right). \allowdisplaybreaks
\end{align*}
We have replaced $f^{'}({\zeta})$ with $\big(f^{'}(\vb*{x}_0)+\vb*{v}\big)$ in the above expressions. Therefore, $O(|\vb*{v}|)=O(\delta)$, and hence we can choose $\delta$ arbitrarily small.
Also, note that $f^{'}(\vb*{x_0})^t(\vb*{x}-\vb*{x_0})=|f^{'}(\vb*{x_0})|H_{\vb*{x}}$ where $H_{\vb*{x}}$ is the height of the point $\vb*{x}$ from the line or hyperplane $AC$, and $\mu(A)=\mu(B)=\frac{1}{2}\mu(\mathcal{L}_n)$. Moreover. if the intersected part of the hyperplane has measure $M$ with respect to $\mu_{p-1}$, then $2M\gamma(\mathcal{L}_n)\geq\mu(\mathcal{L}_n)$ implying that $\frac{1}{\mu(A)}\geq\frac{1}{M\gamma(\mathcal{L}_n)}$. Therefore,
\begin{equation}
    \bar{f}_1-\bar{f}_2= \frac{2|f^{'}(\vb*{x_0})|}{\mu(\mathcal{L}_n)}\left(\int\limits_AH_{\vb*{x}}d\vb*{x}+\int\limits_BH_{\vb*{x}}d\vb*{x}\right) \geq \frac{|f^{'}(\vb*{x_0})|}{M\gamma(\mathcal{L}_n)}\left(\int\limits_AH_{\vb*{x}}d\vb*{x}+\int\limits_BH_{\vb*{x}}d\vb*{x}\right).
\end{equation}
The above integrals are larger than the integrals restricted to the quadrilateral or polyhedron, and the integrated height of the polyhedron is given by $\frac{H^2M}{p(p+1)}$. Hence, we have
\begin{align*}
       & \frac{n_1n_2}{|\mathcal{L}_n|^2}\Big(\bar{w}_1-\bar{w}_2\Big)^2 +  O(|\vb*{v}|) + O\bigg(\frac{\log\log|\mathcal{L}_n|}{\sqrt{|\mathcal{L}_n|}}\bigg)
      > \frac{|f^{'}(\vb*{x_0})|^2} {4\gamma(\mathcal{L}_n)^2p^2(p+1)^2}(H_A^2+H_B^2)^2 \allowdisplaybreaks\\
      & \geq\frac{|f^{'}(\vb*{x_0})|^2}{\gamma(\mathcal{L}_n)^2p^2(p+1)^2}\frac{(H_A+H_B)^{4}}{16}
      = K_p^{-1}\gamma(\mathcal{L}_n)^{-2}|f^{'}(\vb*{x_0})|^{-2}\Big(f^{'}(\vb*{x_0}))^t\overset{\rightarrow}{\gamma_{\vb*{u}}}(\mathcal{L}_n)\Big)^{4}.
 \end{align*}
In the above expressions, $\overset{\rightarrow}{\gamma_{\vb*{u}}}(\mathcal{L}_n)$ represents BD as shown in the above diagram, that is, the vector corresponding to the largest distance along the direction $\vb*{u}$. As $\Delta\leq r_n$, we get
\begin{align}
    \Big(f^{'}(\vb*{x_0}))^t\overset{\rightarrow}{\gamma_{\vb*{u}}}(\mathcal{L}_n)\Big)^{4} &\leq K_p r^{'}_n \gamma(\mathcal{L}_n)^2 |f^{'}(\vb*{x_0})|^{2} 
    \notag\\
    \Big(f^{'}(\vb*{x_0}))^t\overset{\rightarrow}{\gamma_{\vb*{u}}}(\mathcal{L}_n)\Big) &= O\Big({\big(r^{'}_n\gamma(\mathcal{L}_n)^2\big)^{\frac{1}{4}}|f^{'}(\vb*{x_0})|^{1/2}}\Big),
\end{align}
where $r^{'}_n=r_n+O(|\vb*{v}|) + O\Big(\frac{\log\log|\mathcal{L}_n|}{\sqrt{|\mathcal{L}_n|}}\Big)$. We can ignore $O(|\vb*{v}|)<\delta$ as we can choose it arbitrarily small.
As ${\gamma(\mathcal{L}_n)^2}\rightarrow0$, we have
\begin{equation}
(f^{'}(\vb*{x_0}))^t\overset{\rightarrow}{\gamma_{\vb*{u}}}(\mathcal{L}_n))=o\Big((r^{'}_n)^{\frac{1}{4}}\Big).
\end{equation}
Since $|\mathcal{L}_n| > \chi^2_1 / r_n$, and $\frac{\log\log|\mathcal{L}_n|}{\sqrt{|\mathcal{L}_n|}}$ is a decreasing function in $|\mathcal{L}_n|$,
\[
o\bigg(\frac{\log\log|\mathcal{L}_n|}{\sqrt{|\mathcal{L}_n|}}\bigg)\leq o\left({\sqrt{r_n{\Big|\log\big|\log r_n}\big|\Big|}}C_{\chi^2_1}\right).
\]
Hence,
\begin{equation}
    (f^{'}(\vb*{x_0}))^t\overset{\rightarrow}{\gamma_{\vb*{u}}}(\mathcal{L}_n)= o\left(\bigg(\sqrt{r_n \Big|\log\big|\log r_n\big|\Big| }C_{\chi^2_1}\bigg)^{1/4}\right).
    \label{eq::convr}
\end{equation}

Note that the order in the RHS of the Eqn. \eqref{eq::convr} is free of $\vb*{u}$. Using this, we have
\begin{align}
     & {\widehat{f}_n(\vb*{x}_0)-f(\vb*{x}_0)} = \frac{\sum\limits_{z \in \mathcal{L}_n^{\vb*{x}_0}}(f(\vb*{z})-f(\vb*{x_0}))}{|\mathcal{L}_n^{\vb*{x}_0}|} + \frac{\sum\limits_{z \in \mathcal{L}_n^{\vb*{x}_0}}\varepsilon(\vb*{z})}{|\mathcal{L}_n^{\vb*{x}_0}|} \notag \allowdisplaybreaks\\
     &=   o\left(\bigg(\sqrt{r_n \Big|\log\big|\log r_n\big|\Big| }C_{\chi^2_1}\bigg)^{1/4}\right) + o\bigg(\frac{\log\log|\mathcal{L}_n^{\vb*{x}_0}|}{\sqrt{|\mathcal{L}_n^{\vb*{x}_0}|}}\bigg) \notag \allowdisplaybreaks\\
     &= o\left(\bigg(\sqrt{r_n \Big|\log\big|\log r_n\big|\Big| }C_{\chi^2_1}\bigg)^{1/4}\right) +  o\left(\bigg(\sqrt{r_n \Big|\log\big|\log r_n\big|\Big| }C_{\chi^2_1}\bigg)\right) \notag \allowdisplaybreaks\\
     &= o\left(\bigg(\sqrt{r_n \Big|\log\big|\log r_n\big|\Big| }C_{\chi^2_1}\bigg)^{1/4}\right).
     \label{eq::convfin}
\end{align}
The above expression provides the convergence rate of $\widehat{f}$ at the points where the first derivative is continuous and is non-zero. If the function $f$ behaves like a constant locally, then the diameter of the leaf-node at those points can have a non-zero limit as $n\rightarrow\infty$. This is due to {\bf Theorem 3.4}.

\end{appendices}

\bibliographystyle{apalike}
\typeout{}
\bibliography{bibliography.bib}
\end{document}